\documentclass[letterpaper, 10 pt, conference]{ieeeconf}  

\IEEEoverridecommandlockouts                              

\usepackage[pdftex]{graphicx}
\usepackage{amsmath}
\usepackage{amsfonts}
\usepackage{graphicx}
\usepackage{mathtools}
\usepackage{caption}
\usepackage{epstopdf}
\usepackage{color}
\usepackage{hyperref}

\usepackage{subfigure}
\usepackage{verbatim}

\newcommand{\norm}[1]{\left\lVert#1\right\rVert}

\title{\LARGE \bf
Data-driven Modelling of Smart Building Ventilation Subsystem
}

\author{Grigore Stamatescu, Iulia Stamatescu, Nicoleta Arghira and Ioana Fagarasan
\thanks{\emph{Data Availability} The data used to support the findings of this study are available from the corresponding author upon request.}
\thanks{The authors are with the Department of Automatic Control and Industrial Informatics, University "Politehnica" of Bucharest, 313 Splaiul Independentei, 060042 Bucharest, Romania {\tt\small grigore.stamatescu@upb.ro}}
}

\begin{document}

\maketitle
\thispagestyle{empty}
\pagestyle{empty}

\begin{abstract}

Considering the advances in building monitoring and control through networks of interconnected devices, effective handling of the associated rich data streams is becoming an important challenge. In many situations the application of conventional system identification or approximate grey-box models, partly theoretic and partly data-driven, is either unfeasible or unsuitable. The paper discusses and illustrates an application of black-box modelling achieved using data mining techniques with the purpose of smart building ventilation subsystem control. We present the implementation and evaluation of a data mining methodology on collected data over one year of operation. The case study is carried out on four air handling units of a modern campus building for preliminary decision support for facility managers. The data processing and learning framework is based on two steps: raw data streams are compressed using the Symbolic Aggregate Approximation method, followed by the resulting segments being input into a Support Vector Machine algorithm. The results are useful for deriving the behaviour of each equipment in various modi of operation and can be built upon for fault detection or energy efficiency applications. Challenges related to online operation within a commercial Building Management System are also discussed as the approach shows promise for deployment.

\end{abstract}

\section{INTRODUCTION}

\subsection{Smart Buildings Context}

Buildings have become major drivers of energy consumption and quality of life challenges in the modern, urbanised, society. As the potential impact of implementing advanced sensing, computing and communication is steadily realised they have also become smart. 
In a technical context we view and define \emph{smartness} by having the building comply to the dual objectives of occupant awareness and energy efficiency, achieved by modelling, simulation and control over the network of field devices and controllers. This leads to increased requirements on the control strategies to balance in an online manner the needs of the building users for comfort with the needs of the building operator for reduced costs. Furthermore dynamic energy pricing and electrical grid balancing constraints impose real-time requirements which are often addressed by means of demand response (DR) schemes. 

Usually a tool allowing knowledge discovery from large databases with transactional information, customer records or other types of structured and unstructured business information, recently data mining methods have started to be applied to measurements coming from the physical world driven by the emergence of Internet of Things (IOT) solutions in industrial and smart city scenarios \cite{Grosswindhager2017226}. This has been driven to a large extent by more efficient electronic components, communication protocols and advanced algorithms for data processing and dissemination. At their core data mining techniques attempt to identify meaningful, non-trivial, patterns in large bodies of data which are subsequently built into validated models and reevaluated periodically to incorporate new information from the underlying process. The models can be purely data-based or enhanced with domain specific expert knowledge. The data sources are either large static databases of historical measurement which are stored on dedicated machines or in a distributed manner in the cloud or dynamic data streams that have to be evaluated online for timely outcomes.

We focus on the application of such advanced data processing techniques in the field of building automation systems (BAS). In particular, as buildings are key driver of energy consumption in the modern society, by using the information improved decision support systems and control algorithms can be implemented to optimise overall building operation. Within a modular BAS structure, the heating, ventilation and air conditioning (HVAC) subsystem plays a dominant role as both the main driver of energy consumption and of the user satisfaction with the working environment, mainly through subjective assessment of indoor thermal conditions. Air handling units (AHU) are used to ventilate the building with the optimal quantity of fresh air, filtering the air and conditioning it, both heating and cooling, within limited margins. AHU control is responsible for enabling the best ratio of outside and recirculated air which is delivered to the indoor thermal zones of the building. In the context of this contribution the focus is on medium to large commercial buildings where the savings potential has already been validated and the application of advanced data processing and control is thus justified in practical deployments.

\subsection{Related Work}

Though academic and industrial research have developed advanced control strategies, in practice most buildings in which some sort of BAS is available, operate on rule-based control or Proportional Integral Derivative (PID) control loops with static schedules and significant oversight from the building operator. Broad overview of data science methods applied to buildings is presented by \cite{MOLINASOLANA2017598}. The main focus is on assisting building professionals in the field of energy management with appropriate decision support tools. The data mining process is described in conjuction with the specific nature of the application going from raw collected data  to preprocessing, processing, modelling, prediction and validation. \cite{en11040772} proposes a two-level load forecasting architecture for both very short term load forecasting: several hours to one day prediction horizon, and short term load forecasting: one day to several weeks energy prediction. Data processing is following a Lambda Architecture which splits the data into 2 layers. Each layer treats data by a different Autoregressive Integrated Moving Average (ARIMA) algorithm, depending on the needed time window with real-time processing for hourly prediction or slower processing for daily prediction.

A paper focused on energy-efficiency improvements leveraging available building-level data for data mining is \cite{YU201633}. The authors list the main predictive tasks in which data mining of large quantities of measurements and contextual information is relevant. These cover: building energy demand prediction, building occupancy and occupant behaviour and fault detection and diagnosis (FDD) for building systems. \cite{AMASYALI20181192} and \cite{SCHMIDT2018742} further argument through broader studies the relevance of data-driven approaches in timely building energy efficiency applications.

Deployment of distributed sensor networks for finer grained spatio-temporal monitoring of indoor conditions is performed by \cite{8003073}. The authors argue that the statistical modelling of the indoor environment as non-parametric Gaussian processes can lead to reliable information that is fed back to the building management system in order to improve the HVAC control. Wireless sensors can be implemented with limited costs as compared to conventional wired sensors and the monitoring architecture can be adjusted dynamically in order to best capture field level information. In \cite{sensors18} a thermal comfort application using collected HVAC IoT data is presented. Building-level benchmarking data sets \cite{Miller2017439} are highly important to assess algorithm performance and produce reproducible outcomes. The authors present a large database of one year data from 507 non-residential building energy meters, mainly from university campuses. A model based predictive control for maintaining thermal comfort in buildings is applied in \cite{en10030321}. The optimal comfort index is achieved by a cost function depending on both occupant comfort and energy cost.

As compared to traditional model-based control (MBC), data-driven control (DDC) represents an emerging field of study which accounts for the need to manage the data deluge produced by dense temporal and spatial monitoring of various systems. A broad survey on the specific nature of DDC and comparison to MBC in various control structures is discussed by \cite{HOU20133}. Within this concept, the steps of data mining and classification for prediction and assessment are seen mainly as acting as a higher level supervisor to field level control loops in the case of tuning control parameters, set-points and providing contextual information which contributes to improved robustness. One good application example as reference for DDC with random forests of regression trees \cite{Jain:2018:DMP:3174275.3127023}. In this case multi-output regression trees are used to represent the system dynamics over the prediction horizon and the control problem is solved in real-time in closed-loop with the physical plant.

Big data analytics for smart city electricity consumption in presented in \cite{en11030683}. The authors use computational intelligence algorithms to model the consumption of eight university buildings. The outcome consists of offline policies to optimise energy usage across the campus. In \cite{AMAYRI201646} a different application is described using decision trees for occupancy estimation in office buildings. Occupancy modelling and estimation is a critical task in smart buildings as the occupancy level and its accurate forecasting directly impact the HVAC conditioning strategy of the building and avoiding wasteful control. Fault and anomaly detection with a rule-based system is described in \cite{PENA2016242}. The main contributions relate to building automated anomaly detection rules with regard to energy efficiency. This is achieved by combining data mining on historical data with expert information about energy efficiency. \cite{NAJEH201875} illustrate the results of the BRIDGE diagnosis strategy on a dedicated building sensor test bed. By considering sensor faults as data deviations, FDD can accurately detect abnormal conditions. FDD for ventilation subsystems is also covered by \cite{8523895} by using a graph-based approach.

\cite{GONZALEZVIDAL2016994} describe in detail the explicit data modelling process for smart building evaluation. A case study is carried out for energy forecasting of a target building using techniques such a Bayesian Regularized Neural Networks and Random Forests. SVM are also considered but provide weaker results in this specific scenario. Finally in \cite{techrep1} SVM is applied for a regression problem where instead of a class label the output of the algorithm consists of a numeric value.

The current paper also builds upon own previous work dedicated to decision support systems for renewable energy campus microgrids \cite{en10010118} and carrying out Model Predictive Control (MPC) for building simulations \cite{7536017}. Earlier work has also included exploratory data analysis from a single building AHU without further analysis and implementation of learning models at a larger scale \cite{8095166}. In this context we have developed the contributions towards better understanding of collected data from smart buildings. Figure 1 summarises this section with regard to the role of data mining for DDC in this scenario. This generic approach is mapped onto our particular scenario as well. Each of the ventilation units implements local control loops which have to comply to setpoints given by the building operator according to occupancy schedules or seasonal adjustments. Without influencing the low-level control we look at input-output data to indirectly characterise the system behaviour and the end goal of improving the control loop parameters and setpoints through a learning framework.

	\begin{figure*}[t]
	\centering
	\captionsetup{singlelinecheck=false, justification=centering}
	\includegraphics[width=\textwidth]{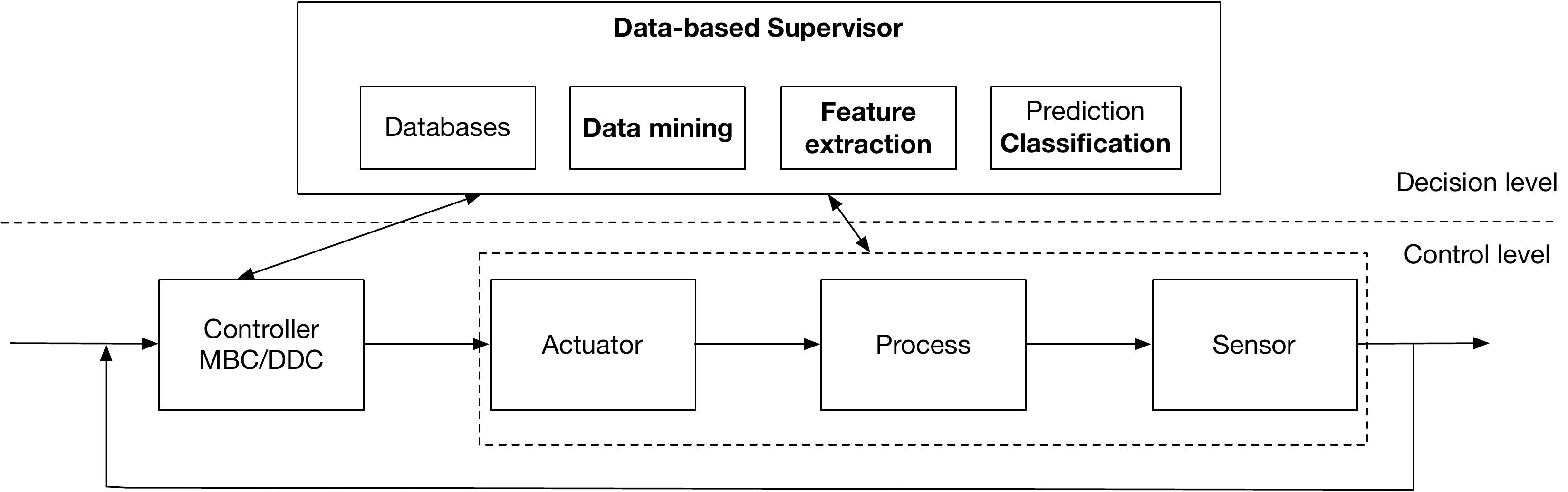}
	\caption{Data supervisory control}
	\label{fig:zero}
	\end{figure*}

\subsection{Objectives and Structure}

The \emph{main objective} of this paper is thus to illustrate the application of a data mining methodology with application to smart buildings for more energy efficient operation. We consider the chosen approach is representative beyond the current case study of learning the behaviour of the ventilation subsystem as well as extensible to other relevant subsystems.

Main contributions of the work can be summarised as follows:
\begin{itemize}
\item argumentation of data-driven building modelling as alternative to conventional system identification or grey-box model approximations;
\item presenting a case study of approaching the black-box modelling of air handling units (AHU) data as a data mining problem with control applications.
\end{itemize}

The rest of the paper is structured as follows. Section II focuses on several specific data mining techniques which can serve as a suitable tool for HVAC subsystem data analysis and processing in smart buildings. Section III reviews the exploratory data analysis steps taken for better in conjunction with expert knowledge from the facility management staff. Section IV presents a case study of one year of multi-AHU collected in a medium size multi-functional campus building. We present here the main results of our study, namely the compressed representation of the time series data, feature engineering and finally results of training various multi-class SVM classifiers on three input datasets. Section V highlights the conclusions of the paper with outlook on future work and implementation.

\section{Data Mining Pipeline: Theoretical Background and Proposed Approach}
	
Time series data-mining \cite{Esling:2012:TDM:2379776.2379788} refers to the application of data mining methodology and algorithms to time series data. When the data is stored in conventional relational databases or unstructured databases, these methods help by representing the data into a new format and allowing the search in the new representation which leads to fewer original data being inspected. Subsequently the method has to offer guarantees about the search results and provide a means to visualise and assess the results. Types of problems that can be solved using time series data mining include similarity matching, clustering and classification of the relevant segments. For classification the segments are determined using methods such as Fourier decomposition, Haar coefficients resulting from wavelet transformations and Piecewise Aggregate Approximation (PAA), becoming features, and then using suitably defined distance metrics the individual examples are assigned to predefined classes. This type of approach has become highly relevant in the Internet of Things age where data collected from the physical world includes a relevant time component whether in the monitoring and control of manufacturing lines, transportation system, the environment and smart city or smart building applications.

We present our approach to apply time series data mining techniques to the problem of building subsystem modelling. The end goal is to extract relevant information from building sensor time series which is can finally assist the facility management department by being integrated into a decision support or directly into the control framework.
The approach illustrates two main techniques, initially we use Symbolic Aggregate Approximation (SAX) \cite{Lin2003,Keogh:2005:HSE:1106326.1106352} to assign symbols to parts of the time series into a unified lower dimension representation. A common technique for classification of discrete patterns of events, Support Vector Machines (SVM) is subsequently illustrated to classify the collected data in order to achieve an indirect model of indoor conditions and operation. The application is focused on the measured exhaust air temperature of the air handling units, as proxy of mixed indoor temperature. We use the other measurement as well as contextual information: outside weather, time of day, scheduling, weekends etc. as additional features in the classification model.

Time series data mining includes various methods that have been applied to non-parametric modelling. SAX has been used by many researchers in various fields and relies on assigning symbols to time series segments based on the observed range of values. Ranges are identified through the data histogram or in a uniform manner. The method provides linear complexity and opens up the use and application of multiple statistical learning tools. One of the tuning factors, the number of regions, can significantly influence the quality of the result.

The method extends PAA \cite{Chakrabarti:2002:LAD:568518.568520} by assigning symbols to the PAA identified segments. The segments can be thus incorporated into a Markov model to compute the probability of the observed patterns for future observations. According to the PAA method description, starting with a time series $X$ of length $n$, this is approximated into a vector $\bar{X}=(\bar{x}_1,...,\bar{x}_M)$ of any length $M\leq n$. Each element of the vector $\bar{x}_i$ is calculated by: 

\begin{equation}
\bar{x}_i=\frac{M}{n}\sum^{(n/M)i}_{j=n/M(i-1)+1}x_j
\end{equation}

This means that we reduce the dimensionality of the time series from $n$ to $M$ samples by initially dividing the original data into $M$ equally sized frame and then compute the mean values for each frame. Putting the mean values together we achieve a new sequence which is considered to be the PAA transform (approximation) of the original data. With regard to computational considerations, the PAA transform complexity can be reduced from $O(NM)$ to $O(Mm)$ with $m$ being the number of frames as tuning parameter of the method. The distance measure is defined as:

\begin{equation}
D_{PAA}(\bar X,\bar Y)=\sqrt{\frac{n}{M}}\sqrt{\sum^M_{i=1}(\bar x_i - \bar y_i)}
\end{equation}

It has been shown by the proposers of the method that PAA satisfies the lower bounding condition and guarantees no false dismissals such that:

\begin{equation}
D_{PAA}(\bar{X},\bar{Y}) \leq D(X,Y)
\end{equation}

Support Vector Machines \cite{esl09} for classification problems extend linear boundary classification problems by imposing a minimum distance $b>0$ from the class separator. In the case of two classes:
\begin{equation}
w \cdot x_k^T + b \geq +1 \quad  \textrm{for} \quad y_k=1 
\end{equation} 
\begin{equation}
w \cdot x_k^T + b \leq -1 \quad \textrm{for} \quad y_k=2
\end{equation} 

SVM searches for the optimal solution by minimising the objective function:

\begin{equation}
J=\norm{w}^2
\end{equation}

which corresponds to maximising the distance $b$. As the classes might be very close to each other or even overlapping, a slack variable $\xi$ is introduced as follows:

\begin{equation}
w \cdot x_k^T + b \geq +1 - \xi_k \quad  \textrm{for} \quad y_k=1 
\end{equation} 
\begin{equation}
w \cdot x_k^T + b \leq -1 + \xi_k  \quad \textrm{for} \quad y_k=2
\end{equation} 

To keep the value of the slack variable $\xi$ small, a penalising factor is added to the cost function:

\begin{equation}
J=\norm{w}^2 + \gamma \cdot \sum_{k=1}^n \xi_k, \quad \gamma > 0
\end{equation}

The SVM paramters $w$ and $b$ are obtained by minimisation of $J$ with the imposed constraints. The vector $w$ represents a linear combination of the training examples as:

\begin{equation}
w = \sum_{y_j=1}\alpha_jx_j - \sum_{y_j=2}\alpha_jx_j
\end{equation}

The general form of SVM is expressed as:

\begin{equation}
\sum_{y_j=1}\alpha_jk(x_j,x_k) - \sum_{y_j=2}\alpha_jk(x_j,x_k) + b \geq +1 - \xi_k, \quad y_k=1 
\end{equation} 
\begin{equation}
\sum_{y_j=1}\alpha_jk(x_j,x_k) - \sum_{y_j=2}\alpha_jk(x_j,x_k) + b \leq -1 + \xi_k, \quad y_k=2
\end{equation} 

where the kernel $k(x_j,x_k)=\phi(x_j)\cdot\phi(x_k)^T$ can take various forms accounting to non-linear class discriminants: linear, polynomial, Gauss, radial basis function, etc. This one significant advantage of SVM, while accounting for increased computing effort and optimisation constraints.

In our classification problem, namely the correct identification of streaming data to one of the installed ventilation units, we encounter a multi-class svm problem, as extension to the binary svm problem described above. This is handled using the one-versus-one approach \cite{KreBel:1999:PCS:299094.299108}. This method leads to the evaluation of all pairs of classifier among the target classes, with $k(k-1)/2$ binary classifiers. The outcome of the classification is achieved by counting the respective votes of each classifier to the test example.

\begin{figure*}[t]
	\centering
	\captionsetup{singlelinecheck=false, justification=centering}
	\includegraphics[width=\textwidth]{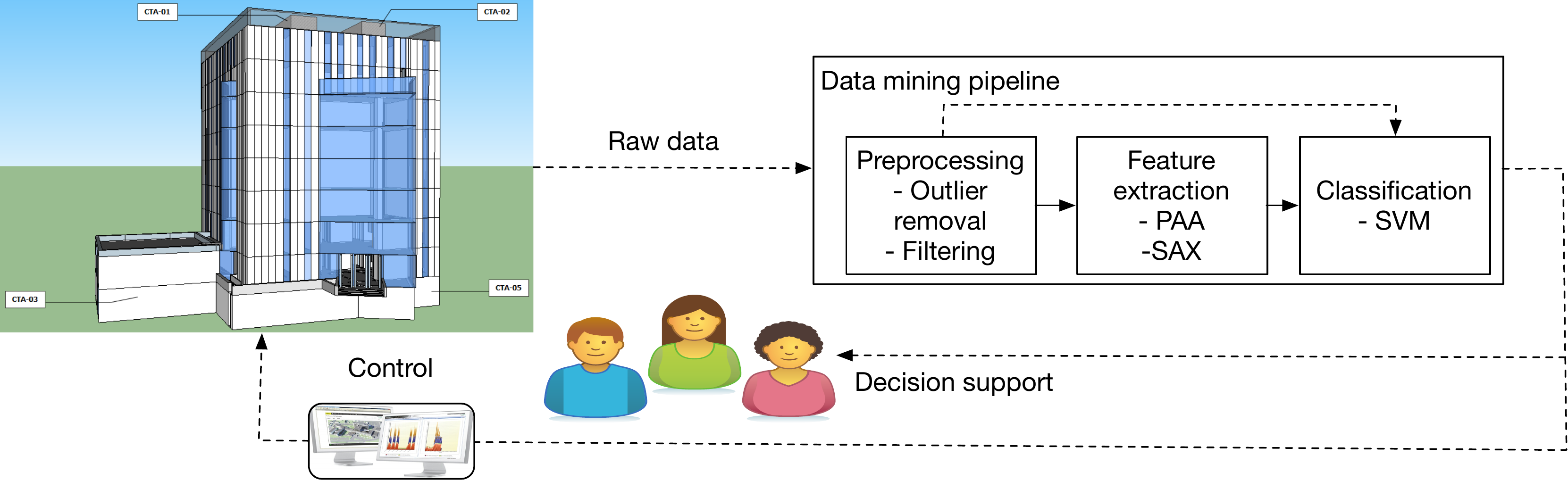}
	\caption{Proposed data mining approach for ventilation subsystem modelling}
	\label{algo}
	\end{figure*}

Figure \ref{algo} summarises the proposed approach with regard to the implemented data processing and learning steps as well as their use either for human decision support our in automatic control loops at the BAS level.

\section{EXPLORATORY DATA ANALYSIS}

We carry out a case study using data mining methodology on measurements taken from four air handling units (AHU) installed in a medium sized campus building. The building is a 7-story, 9000 sqm, facility commissioned in 2016 hosting the PRECIS research center. It contains multiple research laboratories, multi-function spaces, meeting rooms as well as and a large auditorium and administrative offices. It is located at \texttt{44 26'06.0"N 26 02'44.0"E} in a temperate continental climate with hot summers and cold winters. Cooling is handled using on-site electric chillers while heating is provided from a district heating network. For monitoring and control of the various building subsystems, a commercial Building Management System (BMS) software solution from Honeywell is implemented on a central server and used by the facility management staff.

The AHU units are the main components of the ventilation subsystem of the building. They handle the ventilation function of the HVAC subsystem by extracting air from the building and inserting a mix of fresh, air quality objective, and recirculated, energy efficiency objective, air into the building. The target building has four AHU units labeled AHU1 and AHU2 for the top-half of the building and AHU3 and AHU5 for the bottom part. AHU setpoints for input temperature and air pressure are set by the building operator given various seasonal, occupancy and usage factors. In our focus building each AHU handles between 10-20 thermal zones from similarly placed and oriented areas. Though the AHU is controlled by local temperature and air pressure PID loops, the resulting exhaust air from the building is influenced by multiple other factors such as zone-level temperature setpoints input by the user, occupancy and activity levels, internal loads which make the high-level analysis relevant and worthwhile in the absence of extensive field sensors for observing the system. 

Figure 3 presents a picture of the target building along with a representative BMS screen for one of the ventilation units. Access to the BMS software is either local or remotely enabled. 

\begin{figure}[h]
\centering
\subfigure[]{
\includegraphics[width=0.22\textwidth]{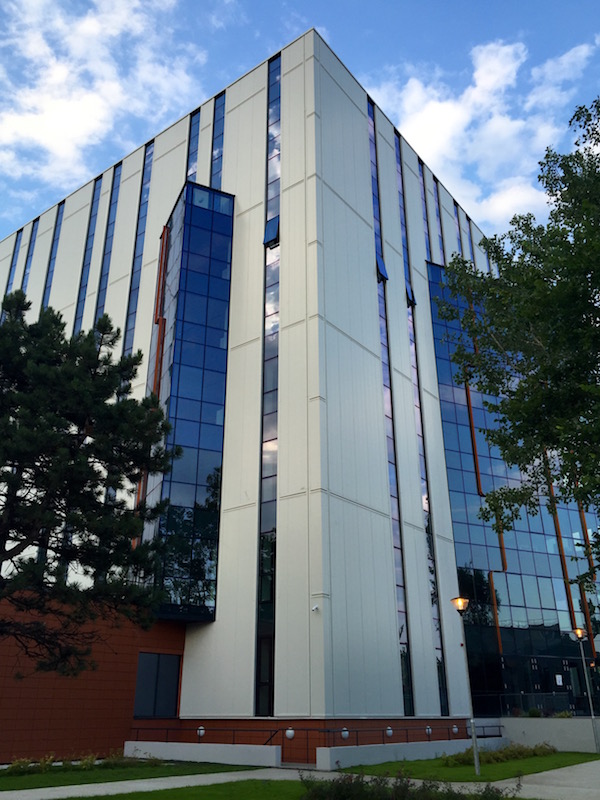}
\label{sfa1}
}
\subfigure[]{
\includegraphics[width=0.45\textwidth]{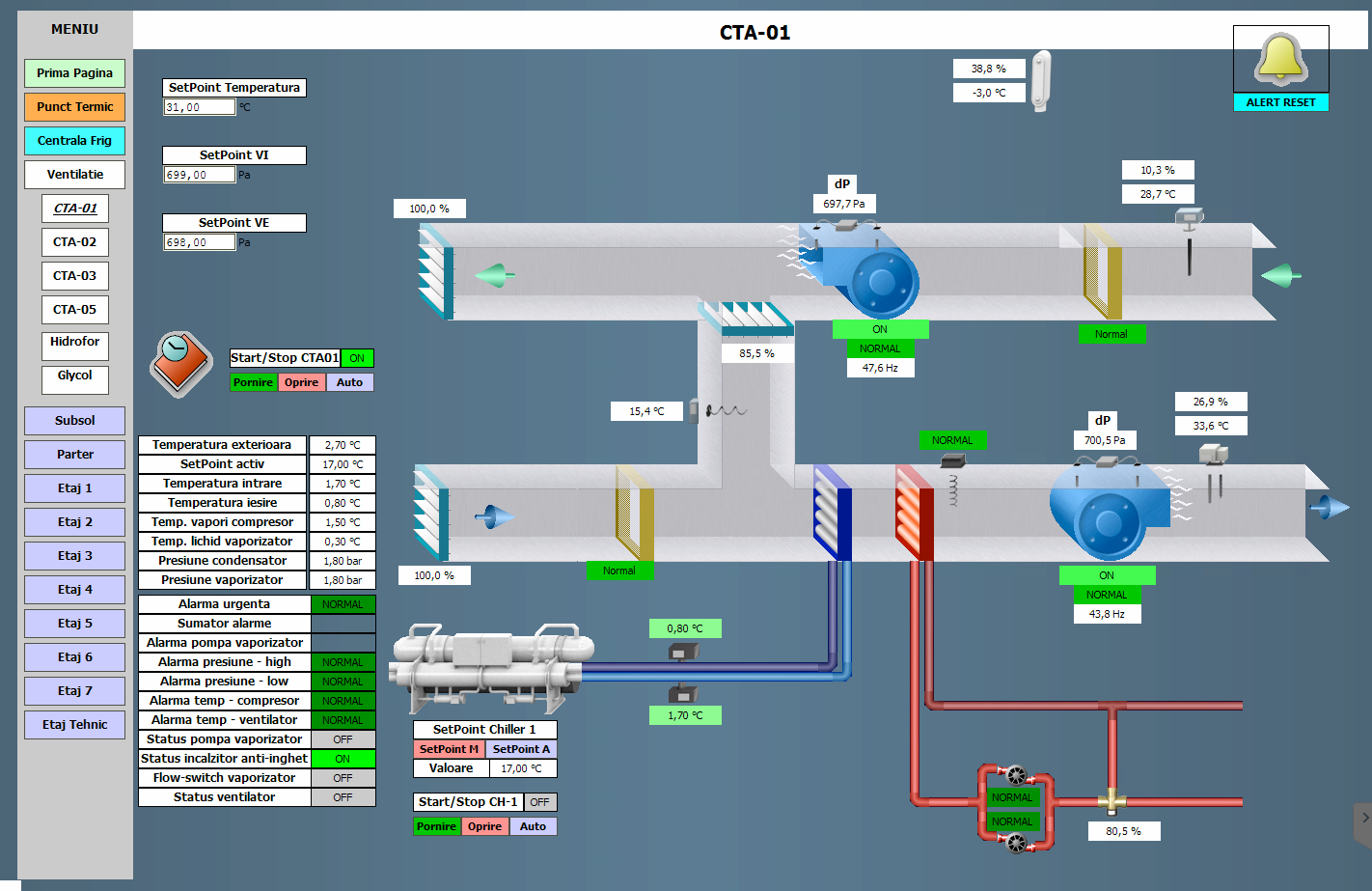}
\label{sfa2}
}
\label{sf2at}
\caption{Target Building and AHU Synoptic Diagram within the BMS Software}
\end{figure}	

Within this technical context it was proceeded to collected the necessary raw data for our study. Data is stored in a SQL structured database on the central BMS server and has been collected offline for the purpose of this study. From each of the four AHU systems we have access to the exhaust, input and recirculation temperatures, sampled at five minute intervals. In addition we collected the reference outdoor temperature and air humidity values from AHU1. Humidity is not used in the current analysis. The reference period of the study is the year 2017, more specific the period from January $7^{th}$ to December $31^{st}$ for a total of 359 days and 103392 data samples for each of the 14 measurement points. The original time series at yearly, monthly and daily timescales are illustrated for AHU1 exhaust and input temperatures in Figure 4. The input temperature refers to the building indoor temperature (AHU output temperature) and the exhaust temperature is determined by the indoor activity disturbance.

\begin{figure}[htb]
\centering
\subfigure[]{
\includegraphics[width=\columnwidth]{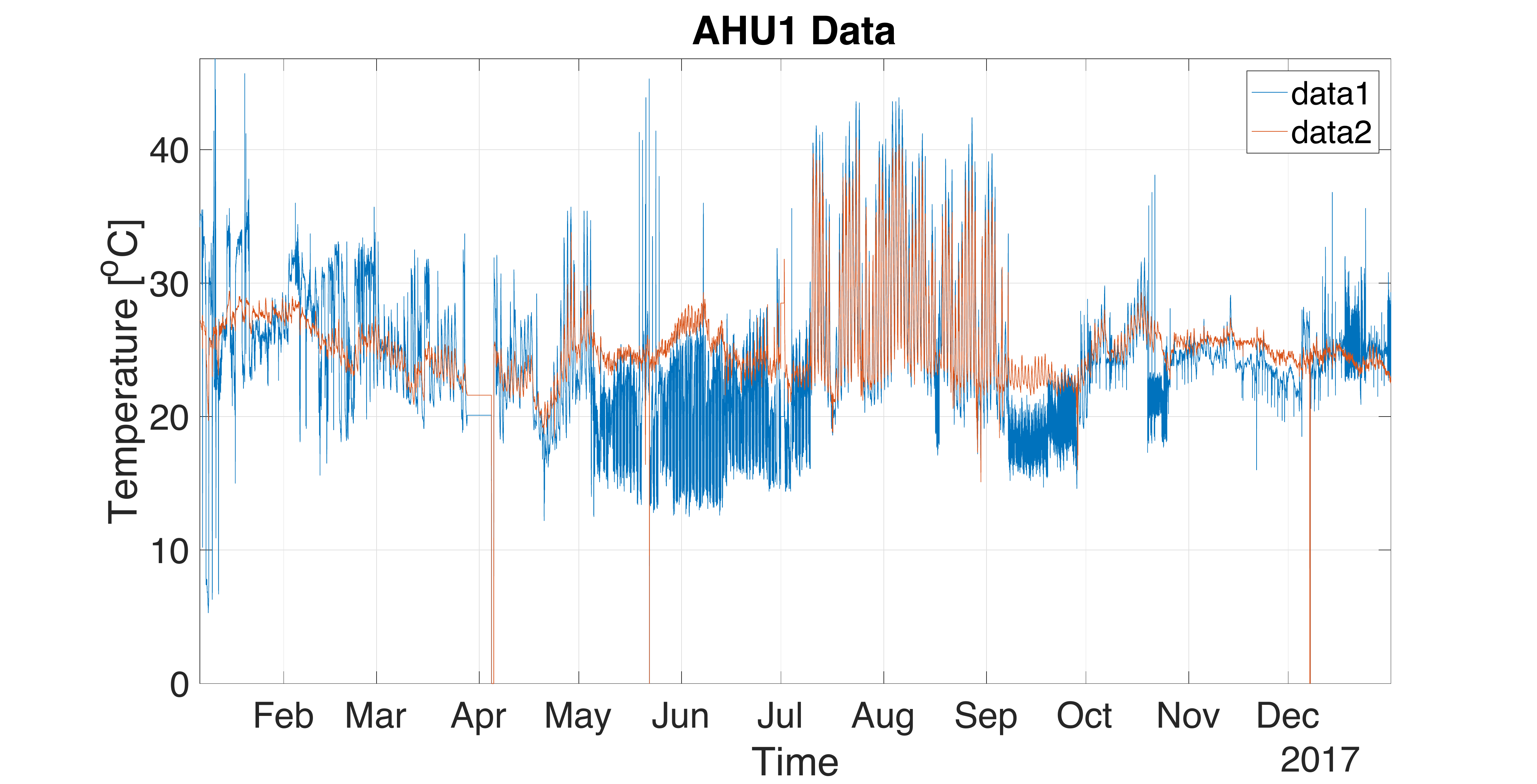}
\label{sf1a}
}
\subfigure[]{
\includegraphics[width=\columnwidth]{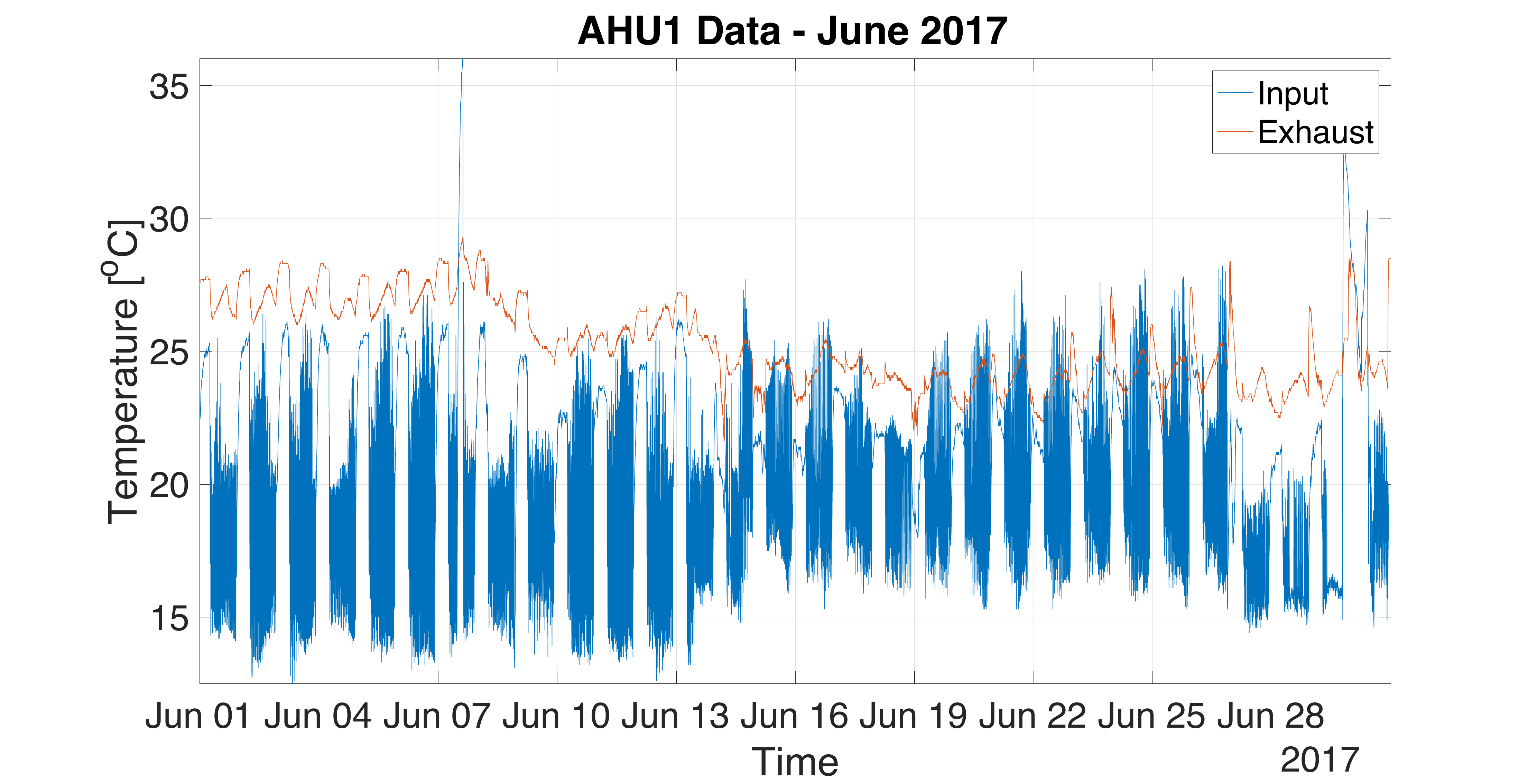}
\label{fig:m2b}
}

\subfigure[]{
\includegraphics[width=\columnwidth]{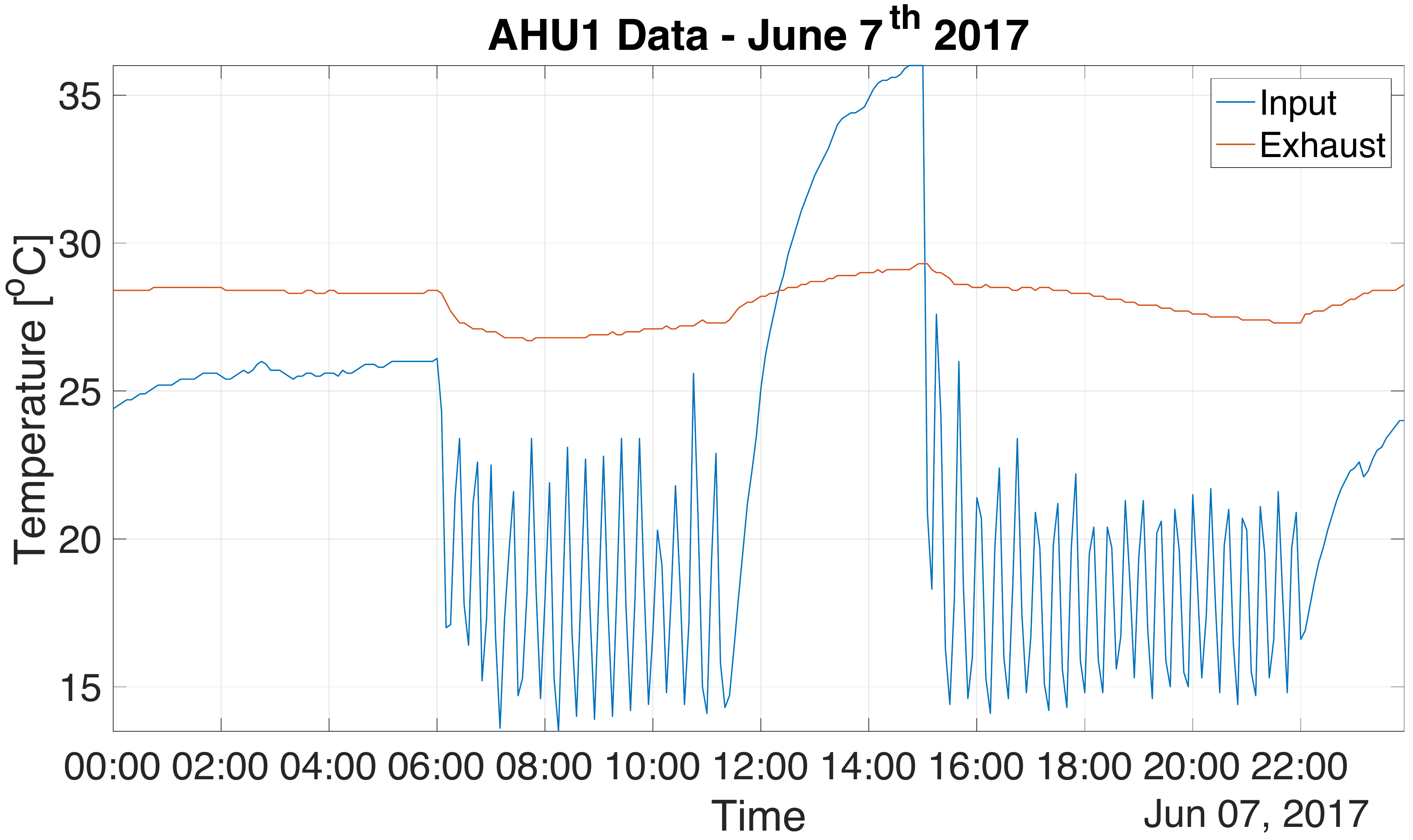}
\label{fig:m2c}
}
\label{sf21}
\caption[Optional caption for list of figures]{AHU1 Input and Exhaust Temperatures a) Yearly b) Monthly c) Daily}
\end{figure}

Preprocessing the raw data allows the improvement of the quality of the data that is input into the algorithms. These steps can help mitigate faulty or erratic sensor behaviour or communication issue that can result in misleading values or gaps in the time series. By analysis the time series we have observed some zero values as well as noise in some of the data.
Two preprocessing steps have thus been implemented:
\begin{itemize}
\item outlier removal - mainly zeros which are replaced with the previous none zero value in the series; the overall occurrence is below $0.1\%$ of the data set;
\item smoothing spline for noise removal even though we sample quite infrequently (5mins); the fitting spline parameters have been adjusted empirically.
\end{itemize}

The resulting time series are input to the next stage in the processing pipeline.

The exhaust temperature data is considered as aggregated proxy for indoor temperature in the corresponding thermal zones assigned to each AHU. The building operator currently sets empirically the input temperature setpoint, pressure setpoint and recirculation ratio setpoint.
The main preprocessing steps taken were to eliminate all zero values, due to sensor faults or communication issues, and using a smoothing spline fit on the time series. We subsequently compute statistical indicators in Table I. These include the minimum, maximum, average and standard deviation, as well as skewness, kurtosis to characterise the underlying probability distribution.

\begin{table}[h]
\label{t3}
\caption{Exhaust temperature statistical indicators}
\begin{center}
  \begin{tabular}{ |c||c|c|c|c||c|}
    \hline
    \textbf{$T$} & \textbf{AHU1} & \textbf{AHU2} & \textbf{AHU3} & \textbf{AHU5} & \textbf{EXT}\\
    \hline
   Min & 15.09 & 16.14 &  0.59 & 9.1 & -15\\
   Max & 40.75 & 38.67 &  30.11 & 32.3 & 39.6\\
   Avg & 25.21 & 23.6 &  21.46 & 25.57 & 13.78\\
   SD & 2.89 & 1.83 &  5.65 & 3.39 & 9.87\\
   \hline
   $s$ & 1.85 & 0.41 &  -2.26 & -1.46 & -0.07\\
   $k$ & 8.51 & 3.74 &  7.64 & 7.04 & 2.46\\
   $r$ & 0.28 & -0.22 &  0.5 & -0.0732 & $--$\\
   \hline

\hline
 \end{tabular}
  \end{center}
  \end{table}
  
  Skewness, $s=E(x-\mu)^3 \setminus \sigma^3$, quantifies the asymmetry of the data around the sample mean. Negative skewness indicates left unbalanced data and positive skewness right unbalanced data. Kurtosis, $k=E(x-\mu)^4 \setminus \sigma^4$, is used as metric for how outlier-prone a distribution is. Distributions which have a value of $k$ higher than 3 are considered to be outlier-prone. In addition we compute Pearson's correlation coefficient $r$ between each measurement array and the outdoor temperature to indirectly determine the recirculation proportion.

Figure 5 presents the histograms for the four data series which illustrate qualitatively the underlying probability distributions for the exhaust temperature values. AHU1-2 values correspond to more predictable zones for laboratory and office space. AHU3-5 cover multi-function and administrative spaces. For subsequent data processing z-score standardisation is applied $x_n=(x-\mu) \setminus \sigma$.

\begin{figure}[h]
\centering
\subfigure[]{
\includegraphics[width=\columnwidth]{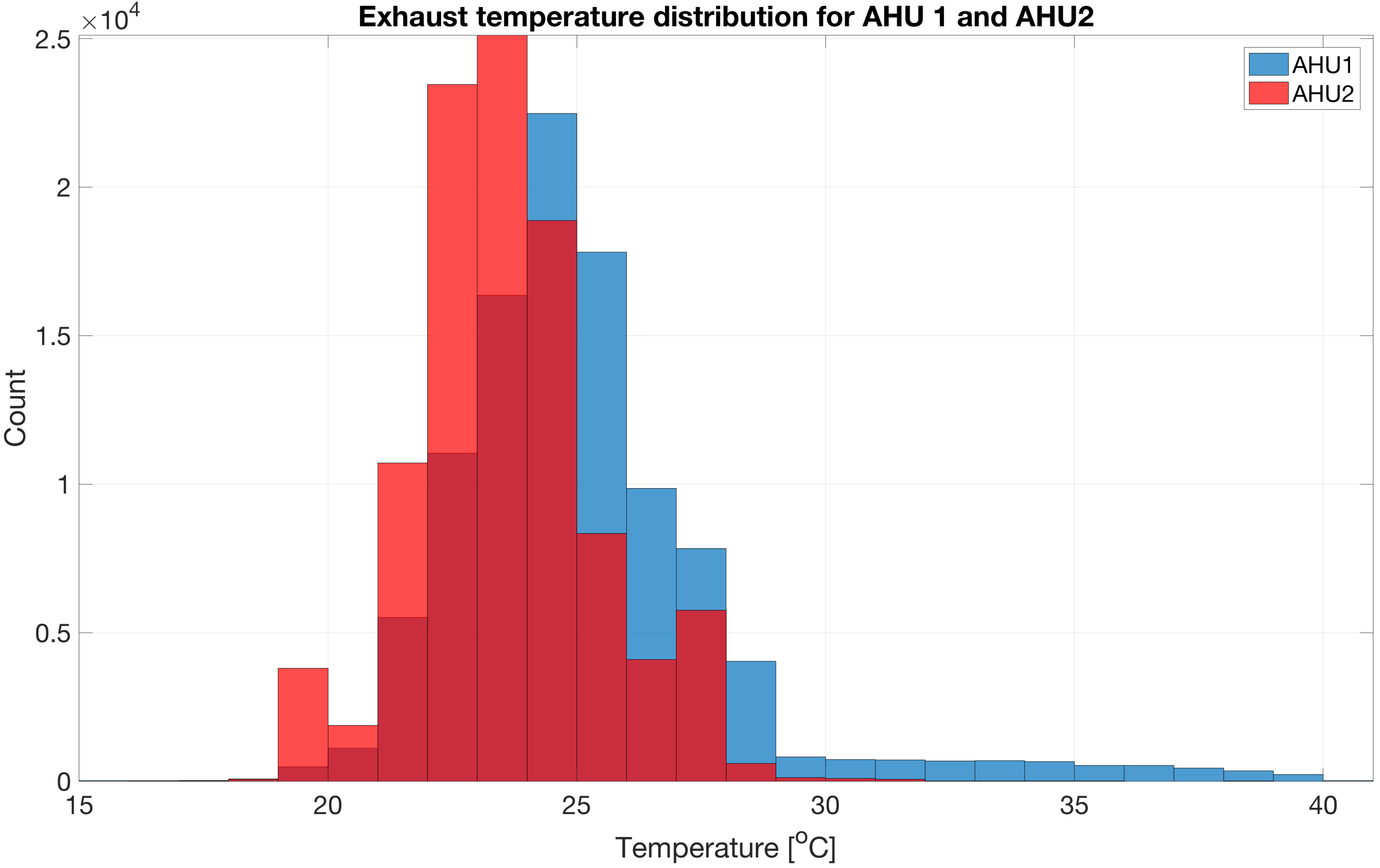}
\label{sf1}
}
\subfigure[]{
\includegraphics[width=\columnwidth]{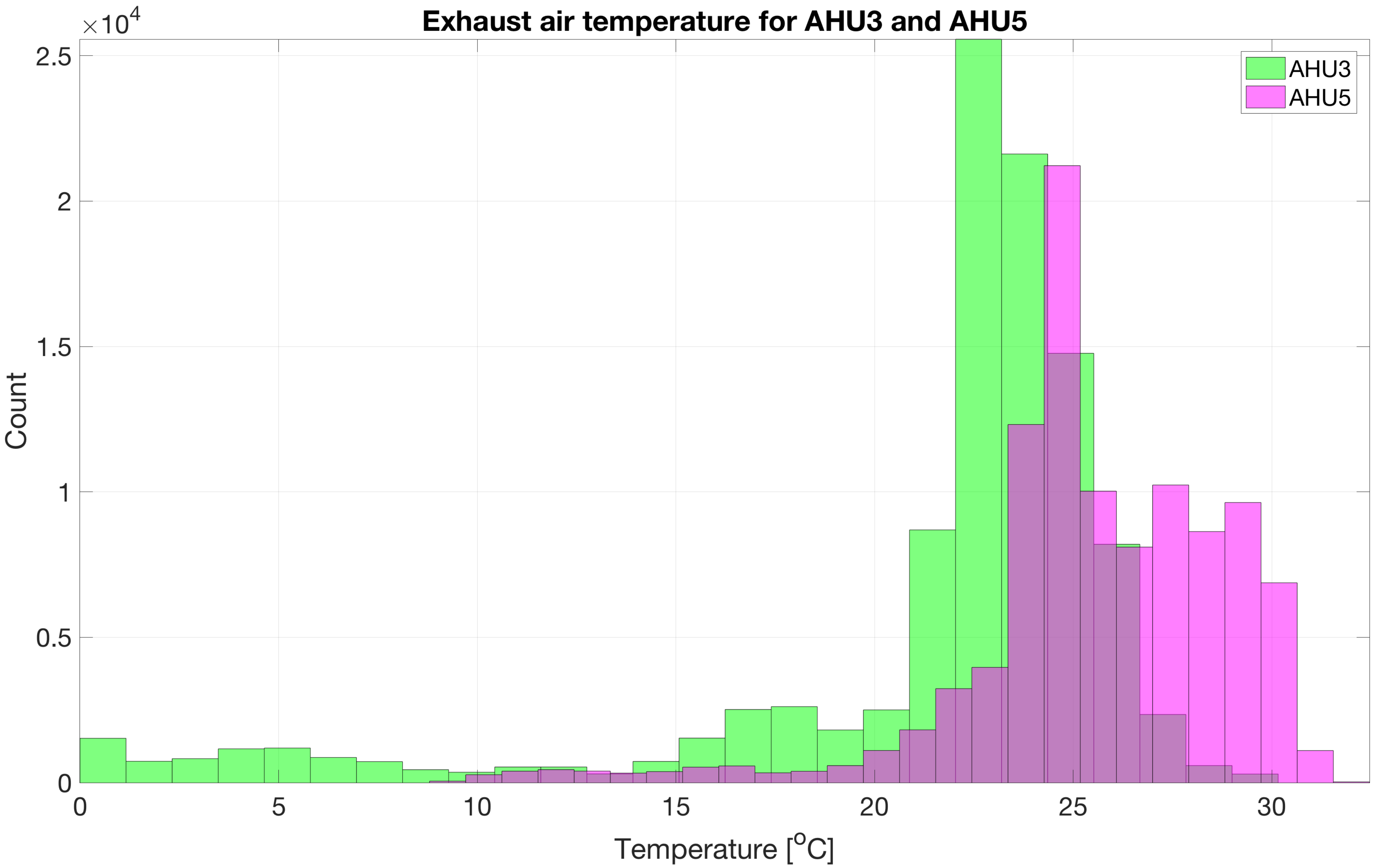}
\label{fig:m2}
}
\label{sf2aaa}
\caption[Optional caption for list of figures]{Exhaust air temperature distributions}
\end{figure}	

As contextual information, Figure 6 illustrates the outdoor temperature for the analysed period. The measurement profile is typical of a temperate continental climate with hot summers and cold winters, thinly separated by shoulder seasons where the best control strategy is no control strategy by leveraging outdoor air as much as possible for ventilation.
	
\begin{figure}[!h]
\centering
\subfigure[]{
\includegraphics[width=\columnwidth]{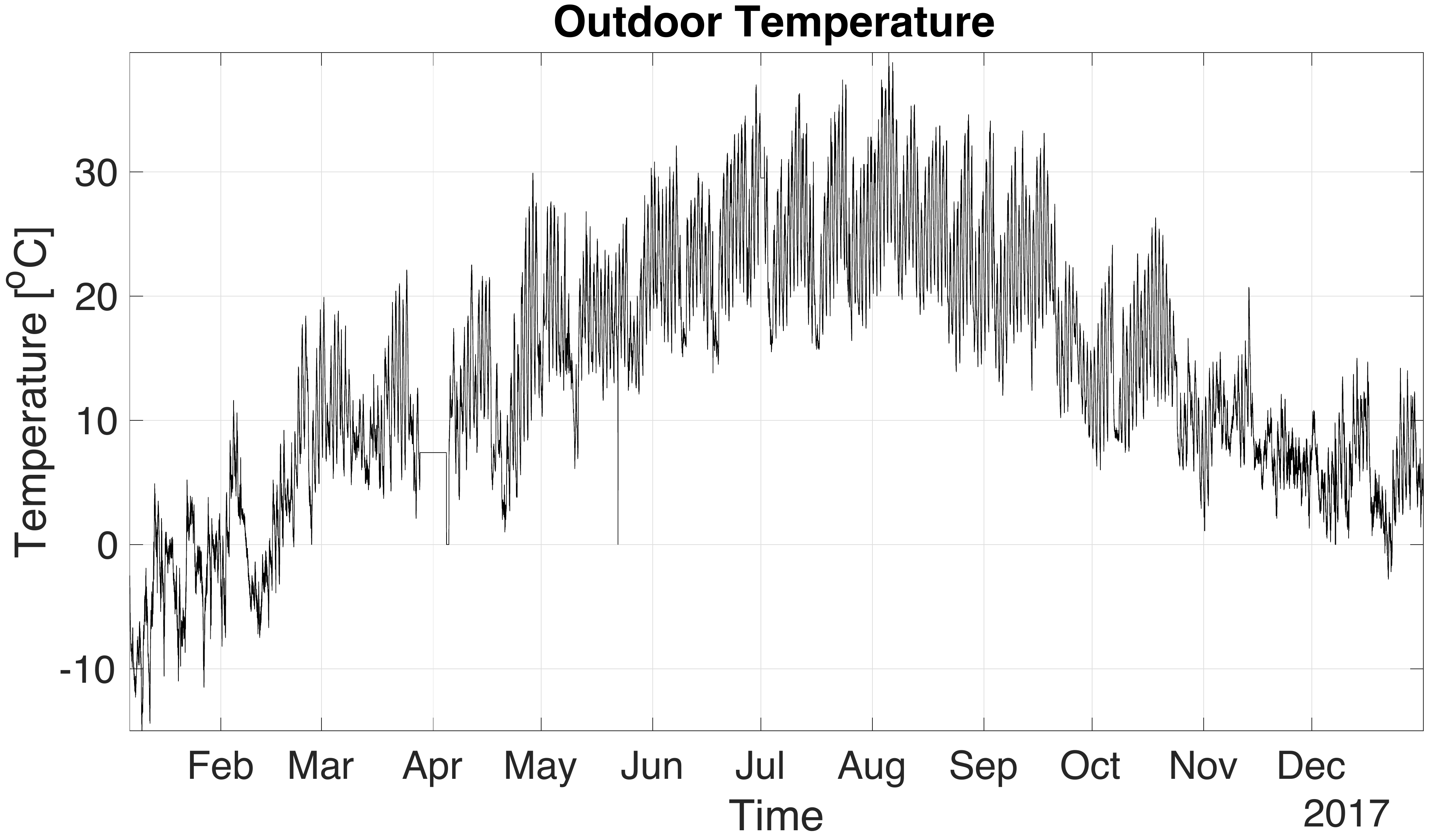}
\label{sf1a}
}
\subfigure[]{
\includegraphics[width=\columnwidth]{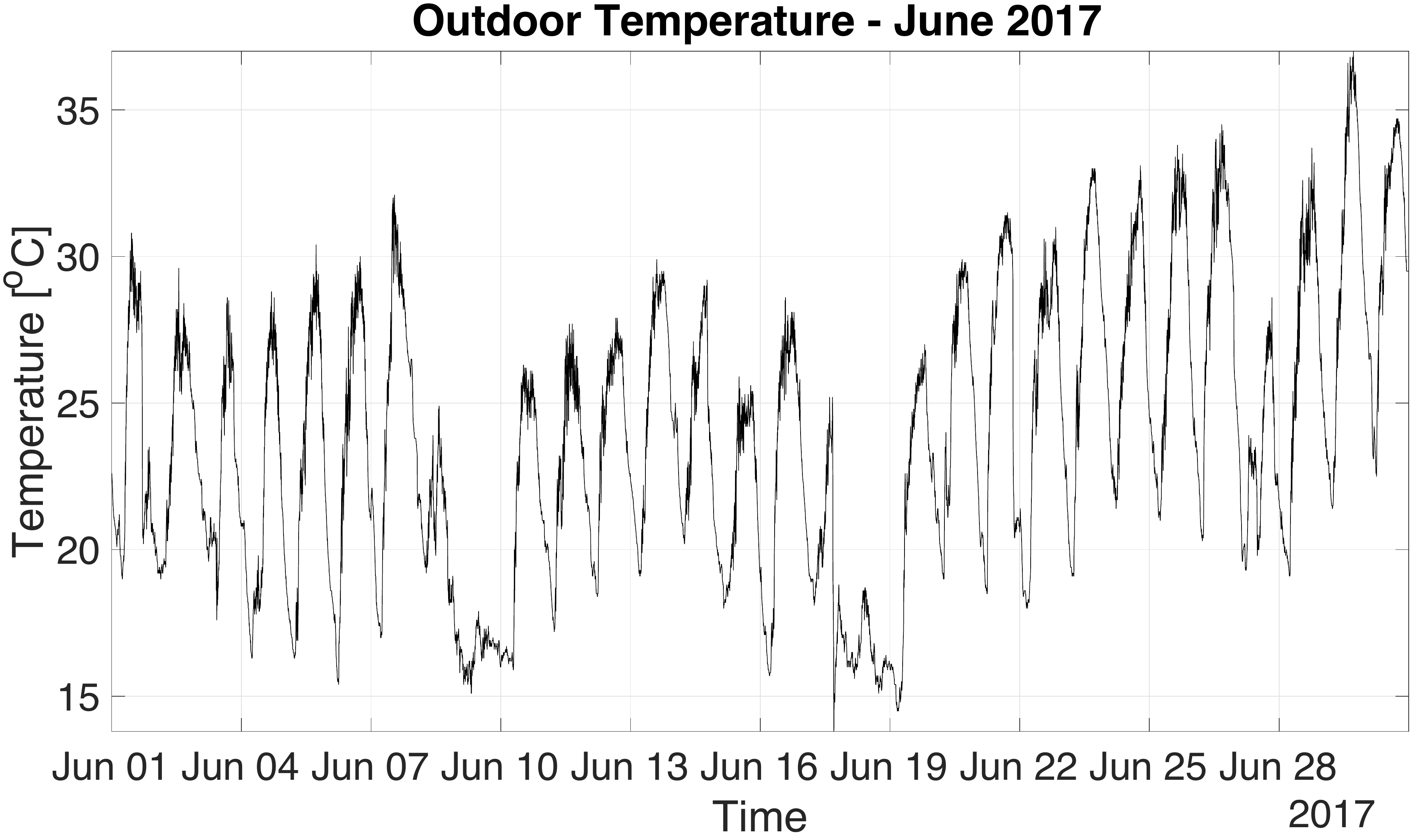}
\label{fig:m2b}
}

\subfigure[]{
\includegraphics[width=\columnwidth]{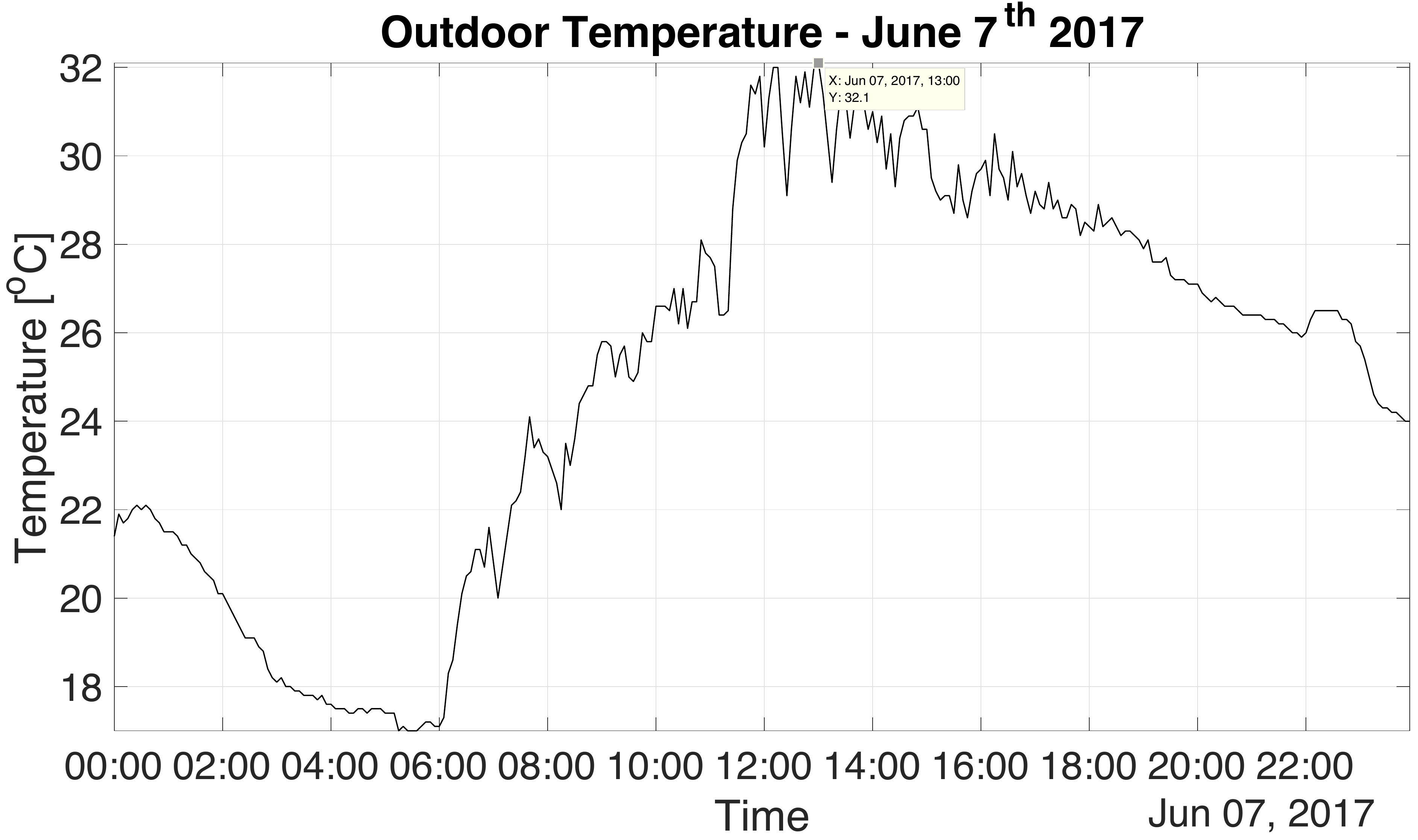}
\label{fig:m2c}
}
\label{zero}
\caption[Optional caption for list of figures]{Outdoor Temperature profiles a) Yearly b) Monthly c) Daily}
\end{figure}

\section{RESULTS}
	
The results section describes the outcome of the following two steps in the data mining processing pipeline. The preprocessed time series are represented into an aggregate form, initially through numeric PAA segments followed by symbolic SAX segments. One of the key tuning parameters is the number of segments for the daily representation of each time series and the alphabet size into which these segments are codified by SAX. A trade-off between the number of approximating elements and granularity of each element is presented. With the chosen configuration this leads to a reduction of a factor of around 30 in the data to be processed. Several SVM classifiers are then trained and tested on the input data with various configurations. For enabling direct comparison, all training is carried out on a single desktop PC, running Windows 7 Professional and MATLAB R2017b, with a quad-core i7 3.6GHz processor and 16GB RAM memory. The results are presented relative to a common benchmark, namely training a svm classifier with a fine gaussian kernel function on the original dataset. We highlight the outcomes while considering that the original dataset contains substantial redundant information given the fine grained timescale of data collection for the slow underlying thermal processes while the aggregated representation might loose some information but provide faster training times which eventually lead to faster online adjustments of an initially lower performing model.

\subsection{DATA AGGREGATION}

We apply the PAA/SAX algorithms to the time series in order to illustrate the patterns derived from each operation mode. By operation mode we intend to better understand the patterns resulting from both the underlying thermal behaviour given occupancy and usage patters, weather fluctuations as well as the control strategy decided for each of the ventilation units by the facility manager. Here we presents the results for AHU1 at different timescales with the input data being sampled at five minute intervals over one year. The time scales are similar to the exploratory data analysis in the sense that the yearly representation allows the observation of major seasonal trends in operation of the ventilation system. At the monthly level we are then able to differentiate better shorter term events and usage and weather variations within each season. Finally, the daily representation is what we use further on as it gives the best representation of usage at a fine grained scale. The tuning parameters of the SAX algorithm are the number of segments $w$ and the alphabet size $a$. The results are shown in Figure 7 at the yearly scale for two configurations: $w=8 \quad a=4$ and $w=10 \quad a=6$. After carrying out multiple experiments on all the available data we find that the latter parameters offer a better and more consistent representation. With the tuning parameters $w=10 \quad a=6$ the monthly and daily sample aggregations for the AHU1 normalised series are illustrated in Figure 8.

\begin{figure}[!h]
\centering
\subfigure[]{
\includegraphics[width=\columnwidth]{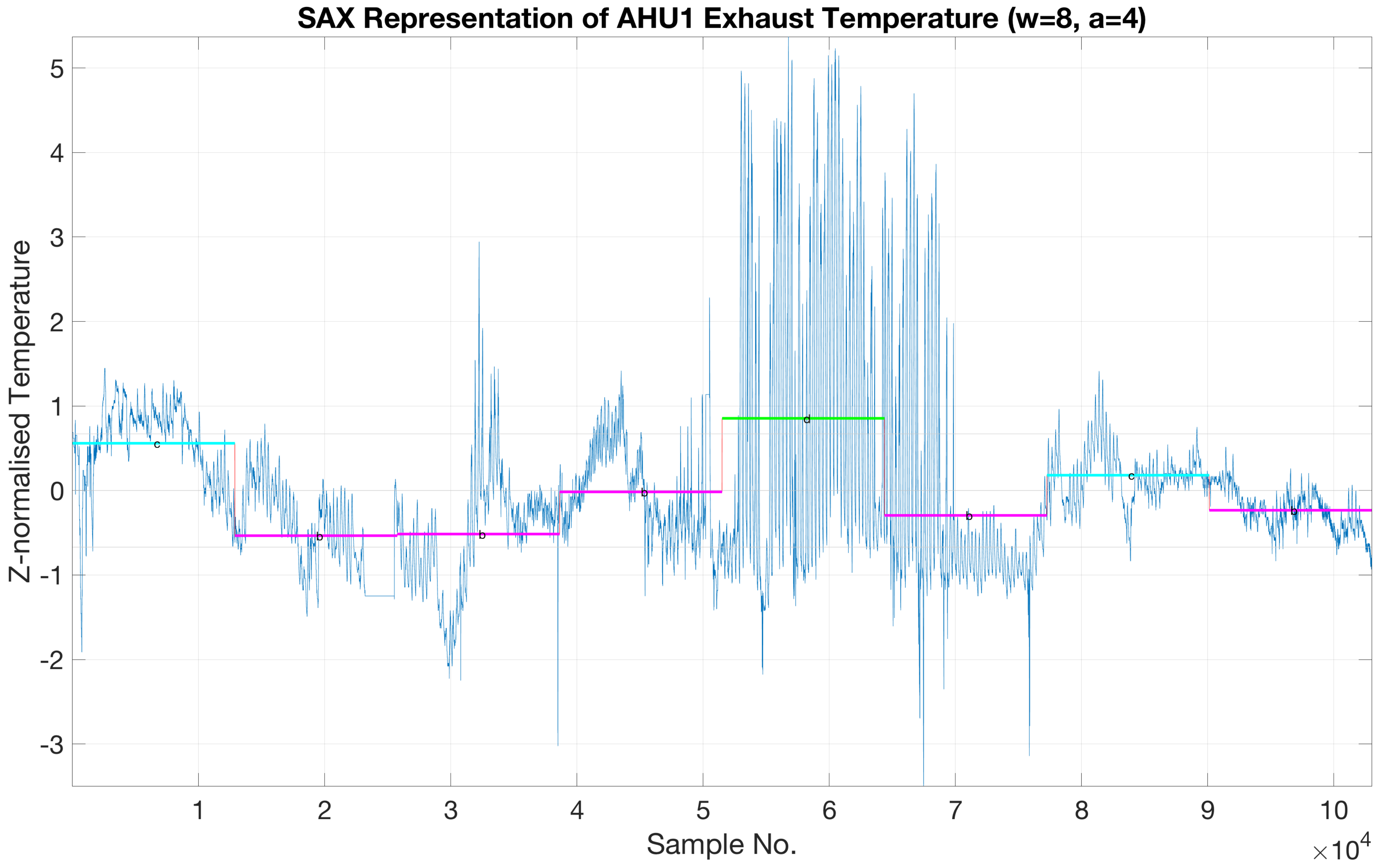}
\label{sf11}
}
\subfigure[]{
\includegraphics[width=\columnwidth]{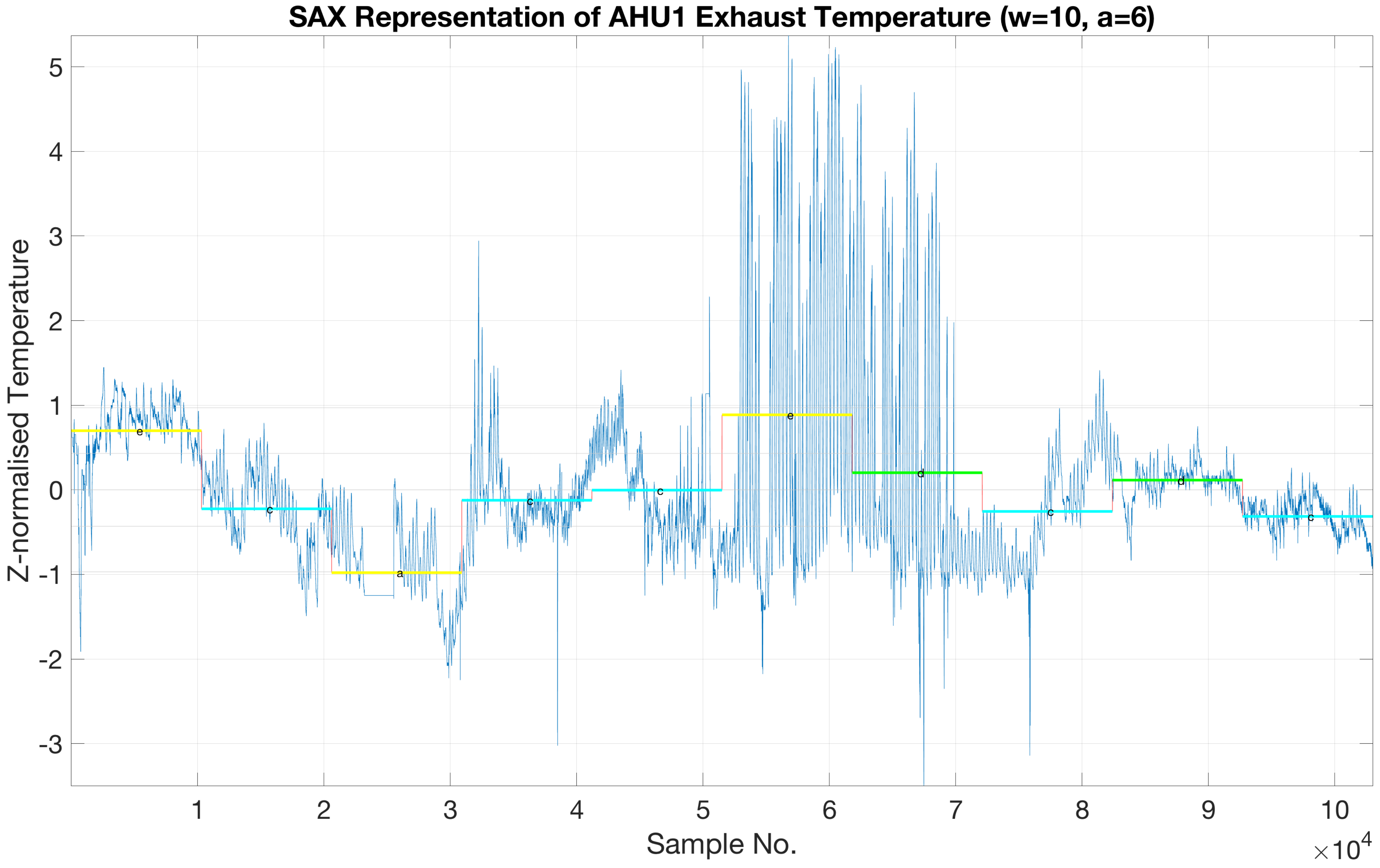}
\label{fig:m21}
}
\label{sf21}
\caption[Optional caption for list of figures]{Yearly SAX Results with different $w$ and $a$ parameters}
\end{figure}	

\begin{figure}[!h]
\centering
\subfigure[]{
\includegraphics[width=\columnwidth]{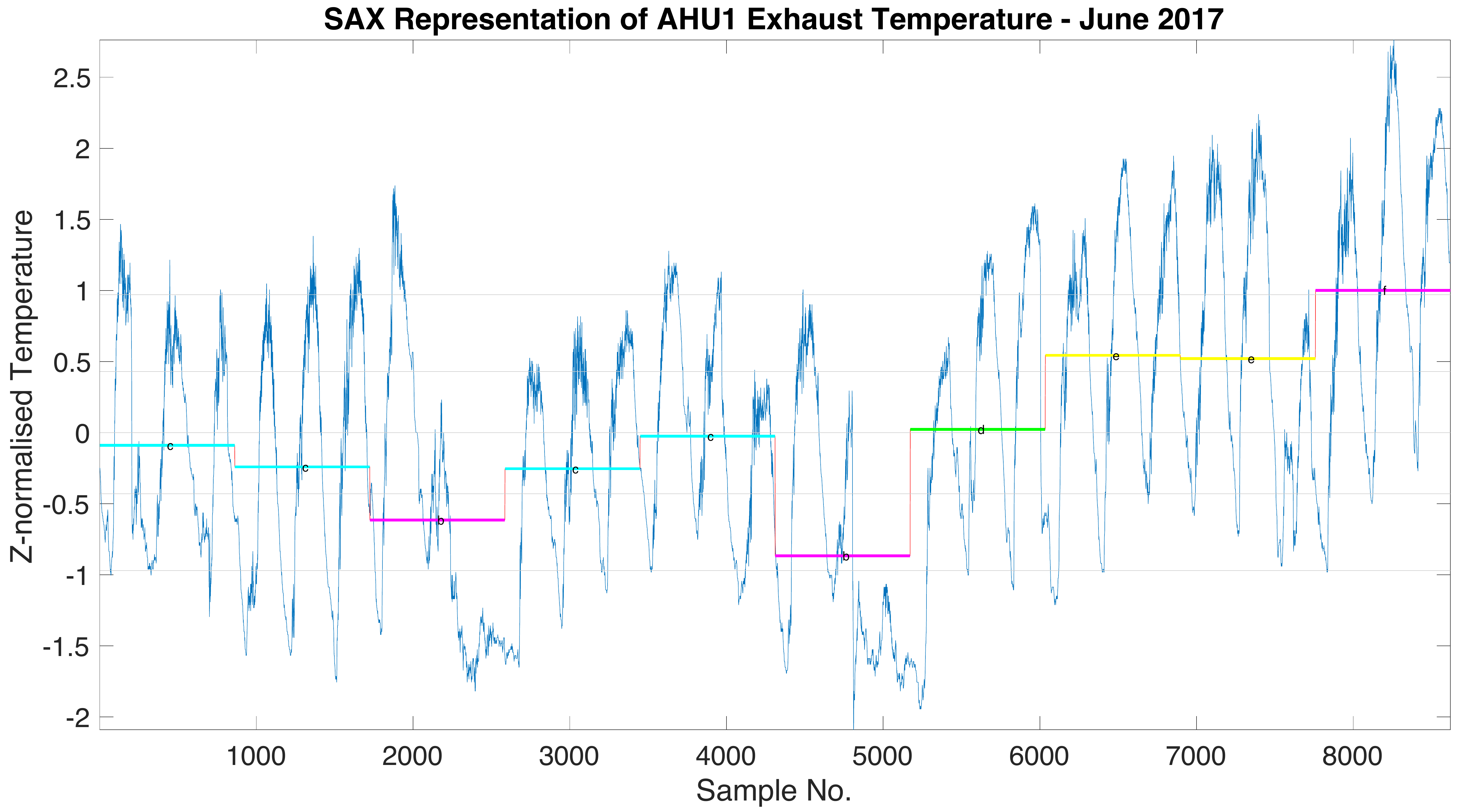}
\label{sf11}
}
\subfigure[]{
\includegraphics[width=\columnwidth]{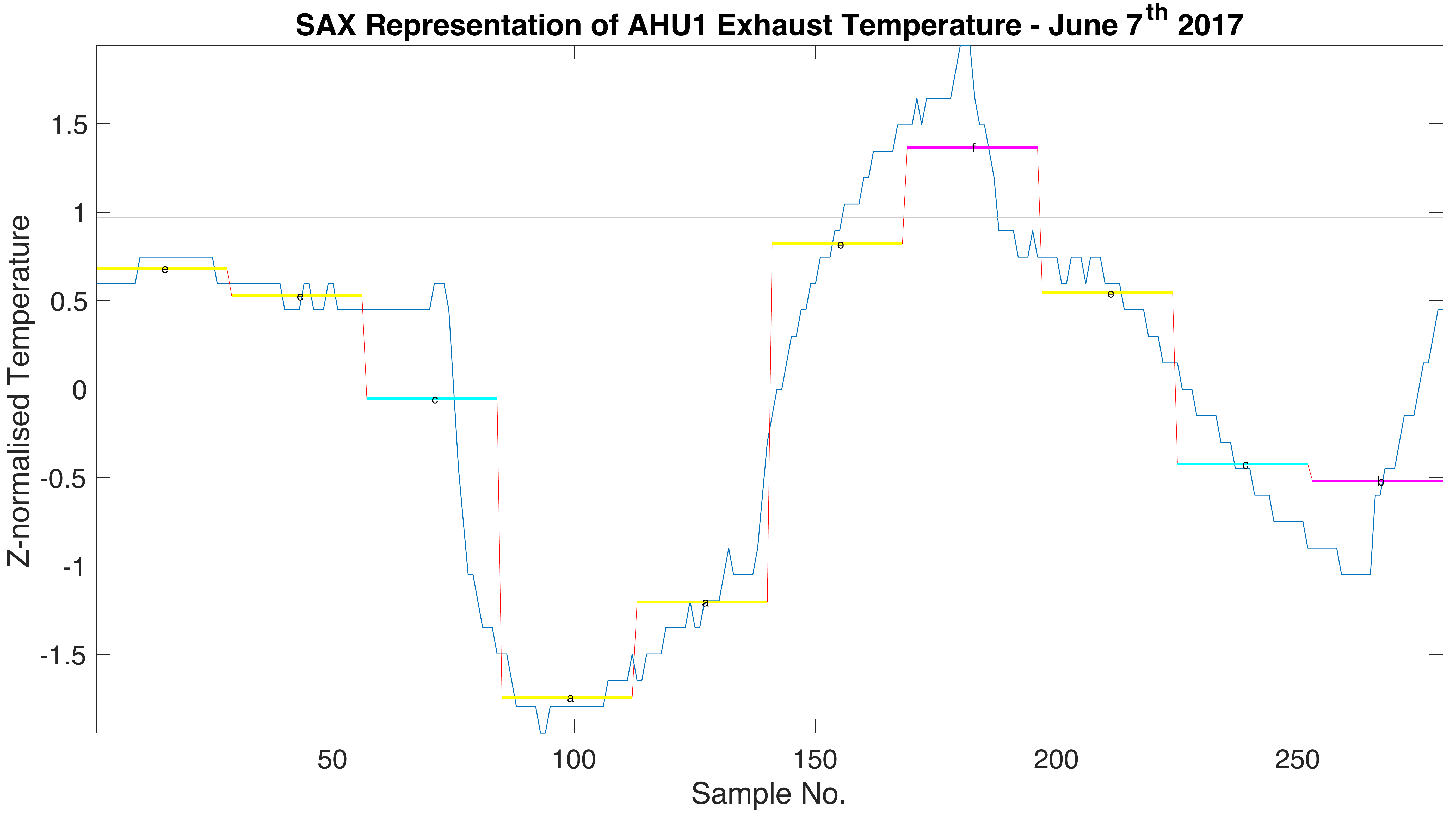}
\label{fig:m21}
}
\label{sf21}
\caption[Optional caption for list of figures]{Monthly and Daily SAX Results with $w=10$ and $a=6$}
\end{figure}	

There are several factors influencing the aggregated patterns which we should account for. For the usage patterns these related to work and school holidays, weekends, structure of the academic year as well as the work schedule at the daily scale. Using the identified sequence we achieve a compact representation of the data, to be correlated with expert knowledge of the building operator. The resulting segments can be used to reconstruct the original data.
Appendix A presents the figures showing full outcome of the application of the SAX method on further data.

\subsection{SVM CLASSIFICATION}

We use SVM because it provides a flexible and powerful classifier when parametrising various kernel functions, in spite of increased computational effort for the optimisation problem. SVM is an inherently binary classifier so for our problem of modelling the patterns of the four ventilation units we use the extension to the multi-class setting as described in Section 3. The one-versus-one approach involves training $k(k-1)/2$ binary classifiers, with $k$ the number of target classes, in our case four, and then voting on the assignment of each observation to a particular class. The advantage is that each of the classifiers are trained on a smaller sample of the input data.
Three datasets result and are used in our experiments for training and investigating the performance of the SVM classifiers: \textit{dataset-proc}, \textit{dataset-paa}, \textit{dataset-sax}. The datasets have the same number of features with different number of observations. \textit{dataset-proc} has 411692 observations representing the preprocessed sensor readings from the ventilation units.  \textit{dataset-paa} and \textit{dataset-sax} both have 14360 observations. For training the classifiers we reconstruct the time series based on the paa and sax representations. The data table header for \textit{dataset-paa} with one example from each class is illustrated in Table 2: 

\begin{table*}[t]
\caption{Dataset PAA header sample}
\centering
\begin{tabular}{cccccccc|c}
\textbf{evac}	& \textbf{in}	& \textbf{rec} & \textbf{ext} & \textbf{iswkn} & \textbf{iswinter} & \textbf{issummer} & \textbf{isshoulder} & \textbf{class}\\
\textit{numeric} & \textit{numeric} & \textit{numeric} &\textit{numeric} & \textit{binary} & \textit{binary} & \textit{binary} & \textit{binary} &\textit{target}\\
0.2501		& 0.4406			& 0.8710 & 0.7291 & 1 & 1 & 0 & 0 & 1\\
0.4426		& 0.8874			& 0.8515 & 0.7291 & 1 & 1 & 0 & 0 & 2\\
1.6611		& 1.4945			& 1.4616 & 0.7291 & 1 & 1 & 0 & 0 &3\\
-1.2388		& -1.1770			& -1.3696 & 0.7291 & 1 & 1 & 0 & 0 &5\\
...		& ...			& ... & ... & ... & ... & ... & ... & ...\\
\end{tabular}
\end{table*}

There are four numeric features originating from temperature measurement at the ahu level and four binary features which codify the temporal aspects with regard to the day of the week and season used in training the models. These are the following:

\begin{itemize}
\item evac -- exhaust temperature of the ahu; normalised temperature values for dataset-raw and dataset-paa and aggregated segment labels for dataset-sax;
\item in -- input temperature of the ahu;
\item rec -- recirculation temperature of the ahu;
\item ext -- outdoor temperature; the same for all units;
\item iswkn -- binary feature used to represent if the measurement was on a weekend (1) or not (0);
\item iswinter -- binary feature used to represent if the measurement was during winter (1) or not (0); 
\item issummer -- binary feature to represent if the measurement was during summer (1) or not (0); 
\item isshoulder -- binary feature used to represent if the measurement was during a shoulder season (1) or not (0); spring and autumn months are considered to be shoulder season.
\end{itemize}

The target class represents the numeric labels of the individual ventilation units 1, 2, 3 and 5.

Based on this input data we go on to train the svm multi-class classifiers and describe the results. As initial testing has shown, the particularities of the data set require a more complex class delimiter so that we present the results obtained with a cubic and a fine gaussian svm respectively. The cubic svm kernel function has the form $k(x_j,x_k)=((x_j)\cdot(x_k)^T+1)^3$ . The gaussian svm kernel function is expressed as $k(x_j,x_k)=exp(-\frac{\norm{x_j-x_k}^2}{2\sigma^2})$. In practice we use a fine gaussian model which uses the kernel scale $\sqrt{P}/4$ with $P$ the number of features. Tuning parameter for the fine gaussian is the scale factor which is determined automatically for the presented results on a similar random seed using a heuristic subsampling approach. A total of six classifiers are thus evaluated, two for each of the three training datasets. Each classifier is evaluated according to the following metrics: accuracy, sensitivity, specificity and area under curve (AUC). All results have being obtained using 5-fold cross-validation for improved statistical robustness.

With regard to classifier evaluation, in the binary case, given the number of positive examples $P$ and negative examples $N$ in the input data set, upon running the classification we obtain a number of true positive $TP$ and true negative $TN$ examples. The accuracy of the classifier is given by $ACC=(TP+TN)/(P+N)$. Sensitivity, also denoted as true positive rate (TPR), is computed as $TPR=TP/P$. Specificity, also known as true negative rate (TNR) is computed as $TNR=TN/N$.  The precision, or positive predicted value (PPV), of a classifier is defined as $PPV=TP/(TP+FP)$. Table 3 shows the aggregated metrics for the six multi-class classifiers.

\begin{table*}[t]
\caption{SVM Training Results}
\centering
\begin{tabular}{c|cc|cc|cc}
SVM 	& Accuracy	& $\bar{AUC}$& Accuracy 	& $\bar{AUC}$ & Accuracy 	& $\bar{AUC}$ \\
 & \textit{dataset-proc}  & & \textit{dataset-paa} & & \textit{dataset-sax} & \\
Cubic	& 81.2 & 0.94 & 63.2 &0.8675 &70.9 & 0.91\\
Gauss & 91.4 & 0.985 & 83.6 & 0.958 &84.6 & 0.965\\
\end{tabular}
\end{table*}

The metrics used for evaluation in our multi-class scenario are defined as follows:

\begin{itemize}
\item Accuracy -- the accuracy metric is a weighted average of the true positive rates across all four classes;
\item AUC -- average area under the curve; as the ROC curve depicts the relation between the false positive rate and the true positive rate when the pairwise discrimination threshold is varied; in our case we report the average AUC between the target class and the sum of the negative classes; 
\end{itemize}

Figures 9--11 below show the best results for each data set in graphical form. We represent the confusion matrix and receiver operating characteristic (ROC) curves for the best AUC indicator.

\begin{figure}[!h]
\centering
\subfigure[]{
\includegraphics[width=0.45\columnwidth]{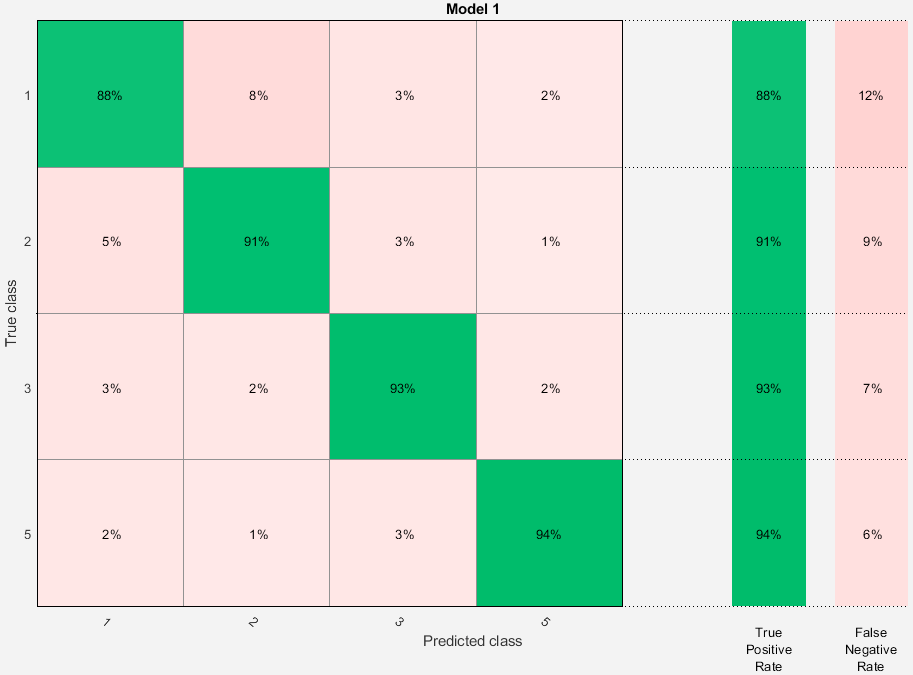}
\label{sf1}
}
\subfigure[]{
\includegraphics[width=0.45\columnwidth]{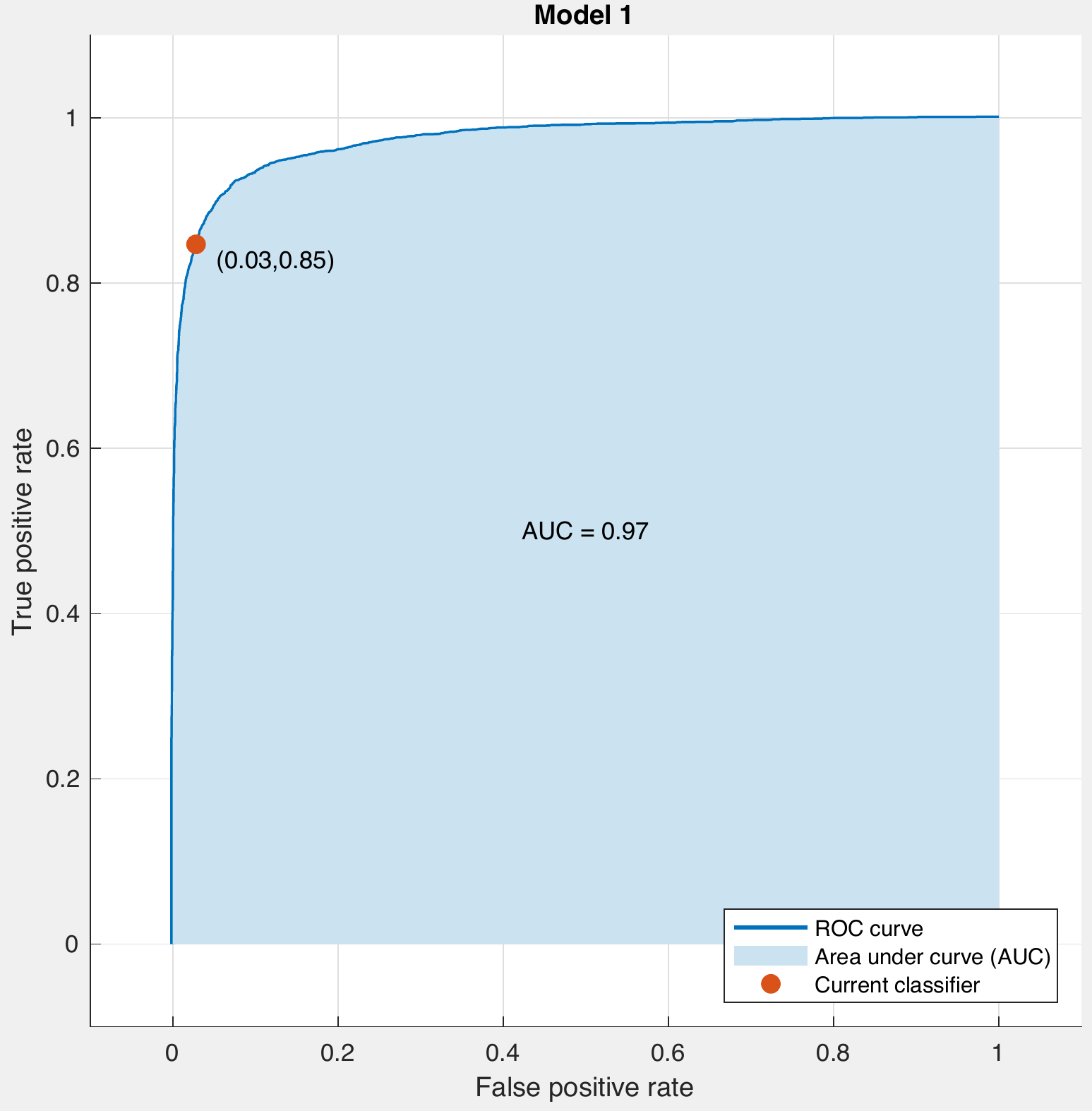}
\label{fig:m2}
}
\label{sf22}
\caption[Optional caption for list of figures]{Dataset-proc SVM performance}
\end{figure}	

\begin{figure}[!h]
\centering
\subfigure[]{
\includegraphics[width=0.45\columnwidth]{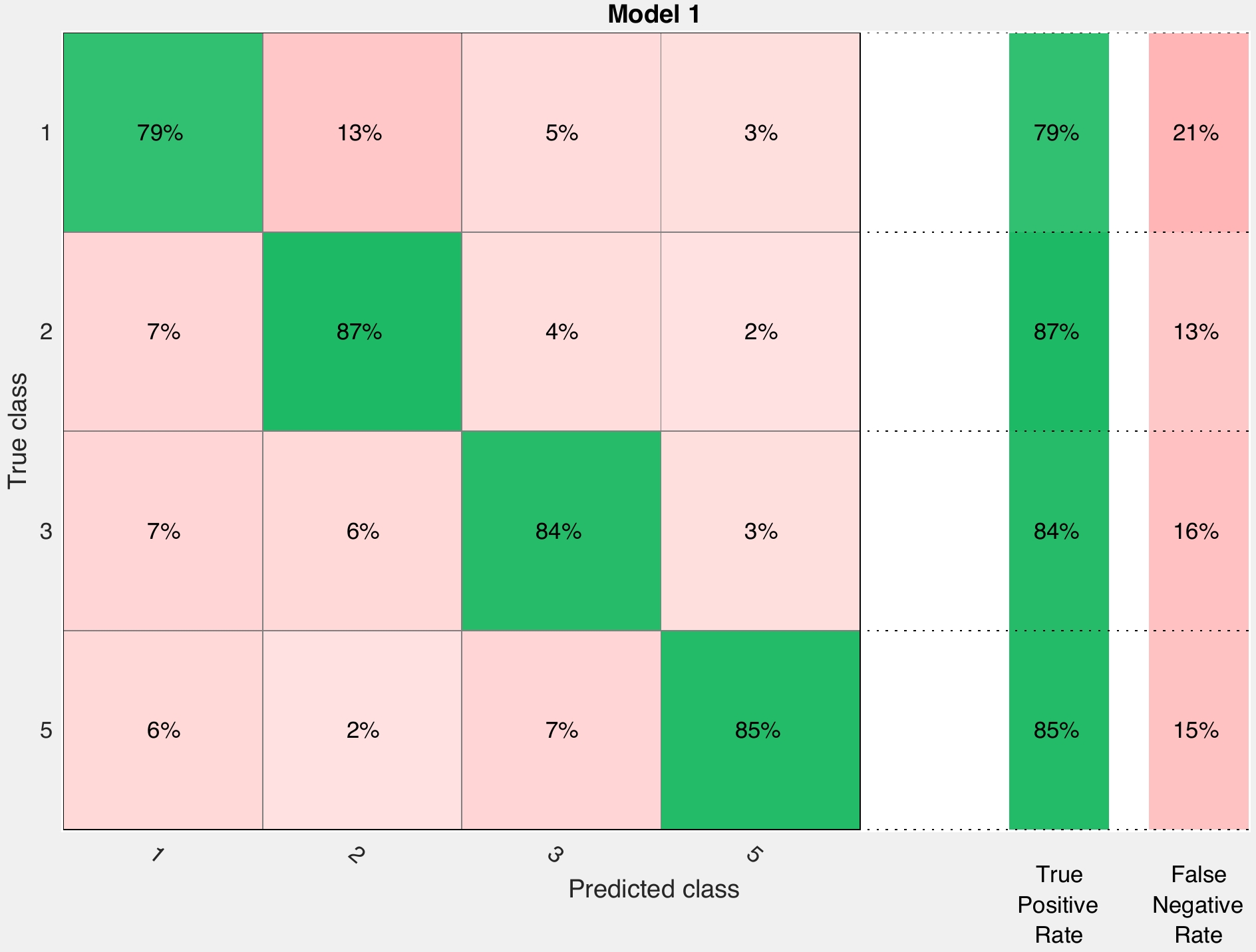}
\label{sf1}
}
\subfigure[]{
\includegraphics[width=0.45\columnwidth]{img/roc_svm_gauss_paa.pdf}
\label{fig:m2}
}
\label{sf22}
\caption[Optional caption for list of figures]{Dataset-paa SVM performance}
\end{figure}	

\begin{figure}[!h]
\centering
\subfigure[]{
\includegraphics[width=0.45\columnwidth]{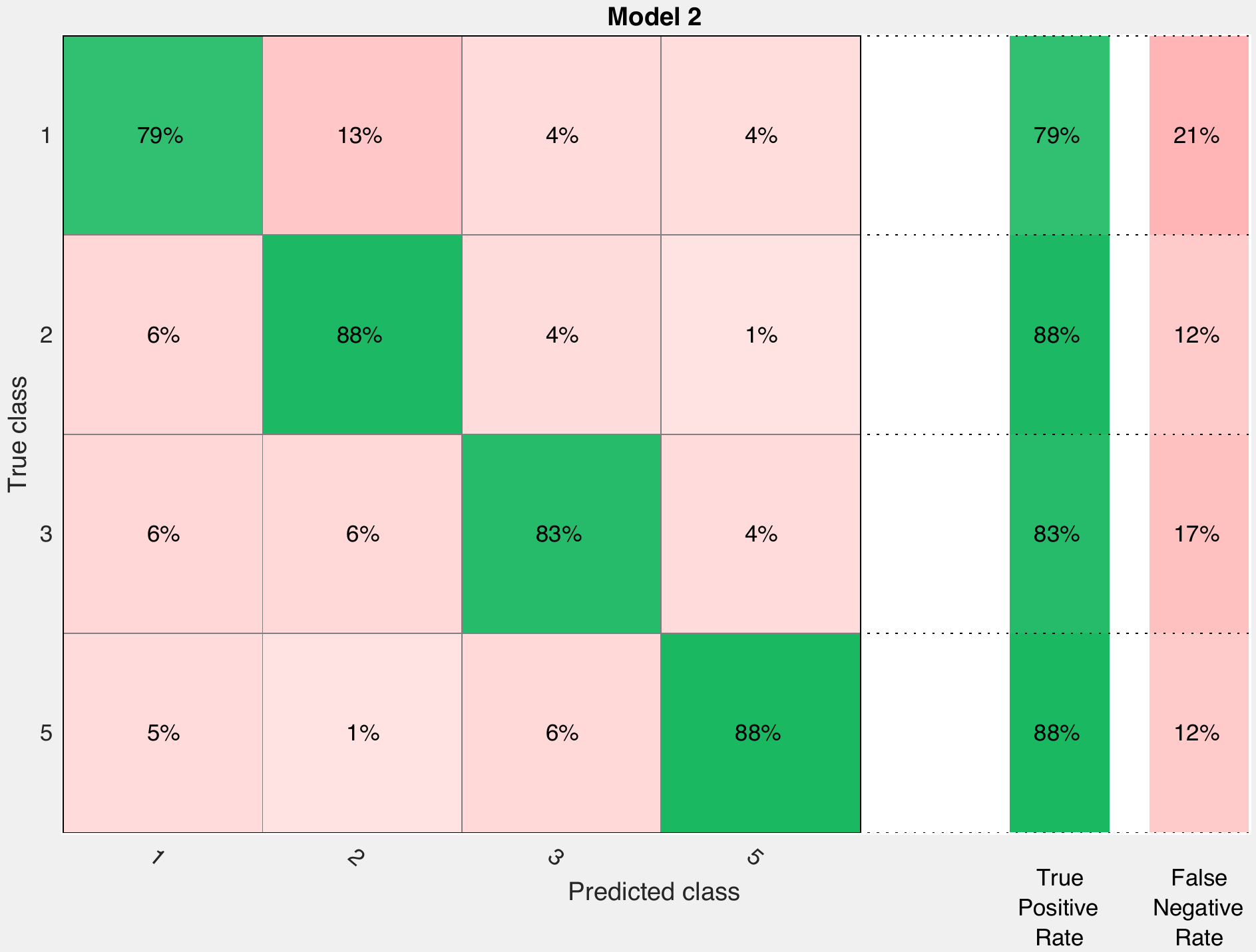}
\label{sf1}
}
\subfigure[]{
\includegraphics[width=0.45\columnwidth]{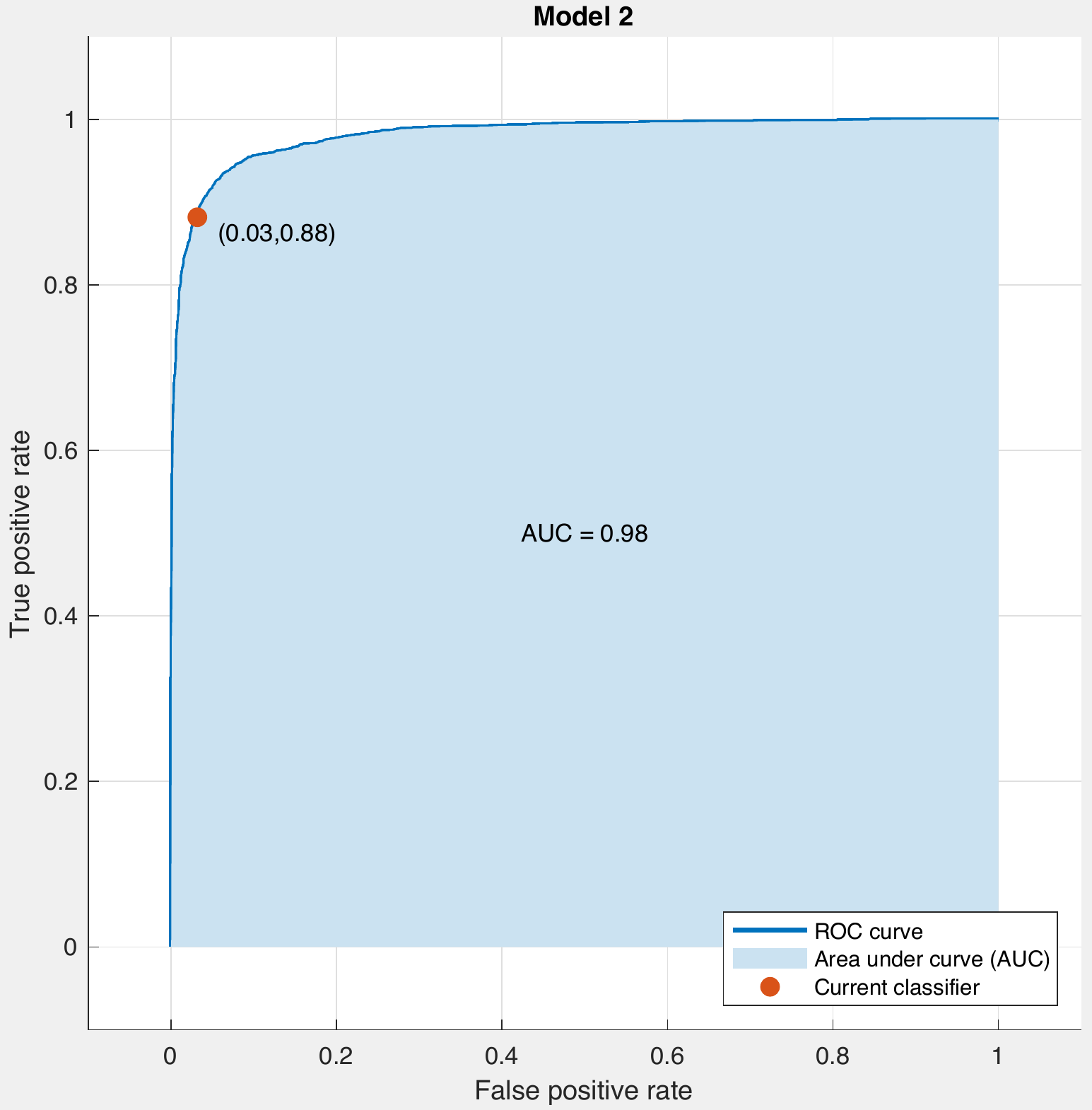}
\label{fig:m2}
}
\label{sf22}
\caption[Optional caption for list of figures]{Dataset-sax SVM performance}
\end{figure}

Finally we present a comparative evaluation of the training time and performance for the case study. Table 4 presents the training time in seconds as well as the prediction speed in observations per second for each of the six trained classifiers.

\begin{table*}[t]
\caption{Performance Analysis}
\centering
\begin{tabular}{c|cc|cc|cc}
SVM 	& Training Time 	& Pred. speed & Training Time 	& Pred. speed & Training Time 	& Pred. speed \\
 & \textit{dataset-proc}  & & \textit{dataset-paa} & & \textit{dataset-sax} & \\
Cubic	& 125321 & 1150 &5136 &9600 &4868 &12000\\
Gauss & 14388 & 450 & 58.172 & 4200 &48.773 & 4600\\
\end{tabular}
\end{table*}

A graphical representation is shown in Figure 12 which is useful to assess the magnitude of the improvements. We observe how, with the same number of features, when the number of observations of the input dataset decreases by a factor of 30 the training time decreases roughly by a factor of 250, validating the nonlinear complexity increase of the svm classifier with regard to input dataset size. An experimental observation is that given the automatic determination of the kernel scale factor used in which is based on a heuristic subsampling method, training times for the cubic polynomial svm kernel are significantly larger than the gaussian kernel svm on the datasets used. 

	\begin{figure}[!h]
	\centering
	\captionsetup{singlelinecheck=false, justification=centering}
	\includegraphics[width=\columnwidth]{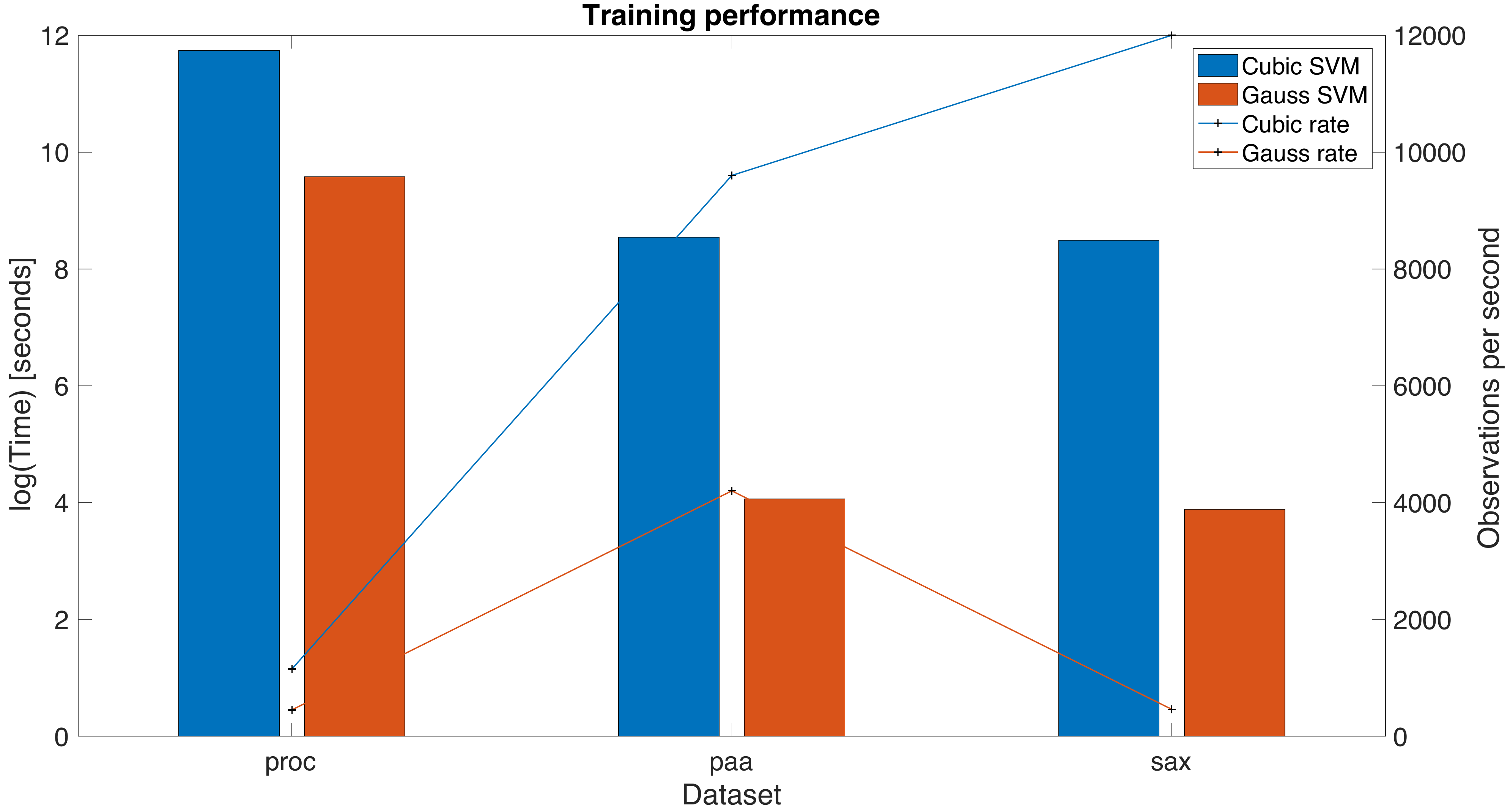}
	\caption{Performance summary}
	\label{fig:zero}
	\end{figure}
	
\section{CONCLUSIONS}
	
The paper has presented a practical application of data mining methodology to data collected from the ventilation subsystem of a smart building. Motivation of the work has been mainly driven by the need to better understand black-box building and operator dynamics in order to improve the control strategies. The presented framework is applicable beyond the described case study to other challenges which might relate to occupant comfort and/or energy efficiency in a building context. The results are replicable through the available Matlab datasets and associated scripts. The important contribution of the study has been the resulting model of the Air Handling Unit of the studied smart building which can be used further with direct impact on the energy efficiency of the ventilation system.

Our main goal has been to achieve a better understanding indoor temperature and ventilation operational patterns which can lead to an improvement in the building control, either off-line or on-line. Future work will be focused on automatic adjustment of AHU setpoints based on learned analytical rules and models. Thus the data hypervisor control would be achieved. For on-line operation, in order to link the developed software routines and modules with the commercial building management system, the implementation of a suitable middleware platform such as VOLTTRON \cite{7725895} in the target building is foreseen. With regard to the core data mining and learning techniques, further investigation into the most relevant kernel functions and hyperparameter tuning can be pursued. Both high level tools and low level libraries for machine learning offer good potential to improve training performance, including by leveraging cloud based infrastructures for data analysis and processing.

\section*{Conflicts of Interest}
The authors declare that there is no conflict of interest regarding the publication of this paper.

\section*{Funding Statement}
The research received no specific funding and was carried our as part of the employment of the authors with the University Politehnica of Bucharest, Romania.


\bibliographystyle{IEEEtran}
\bibliography{IEEEabrv,sensors_new}

\end{document}